\documentclass[journal]{IEEEtran}
\usepackage{amsmath,amsfonts}
\usepackage{algorithmic}
\usepackage{algorithm}
\usepackage{array}
\usepackage{amssymb}
\usepackage{xhfill}
\usepackage{graphicx}
\usepackage{epstopdf}
\usepackage{float}
\usepackage{textcomp}
\usepackage{stfloats}
\usepackage{url}
\usepackage{threeparttable}
\usepackage{amsthm}
\usepackage{pifont}
\usepackage{cite}
\usepackage{verbatim}
\usepackage{colortbl}
\usepackage{makecell}
\def\BibTeX{{\rm B\kern-.05em{\sc i\kern-.025em b}\kern-.08em
    T\kern-.1667em\lower.7ex\hbox{E}\kern-.125emX}}
\usepackage{balance}
\usepackage[caption=false,font=footnotesize,labelfont=rm,textfont=rm]{subfig}
\usepackage[colorlinks,linkcolor=blue,anchorcolor=blue,citecolor=blue,urlcolor=blue]{hyperref}

\newtheorem{theorem}{Theorem}
\newtheorem{corollary}{Corollary}
\newtheorem{lemma}{Lemma}
\newtheorem{remark}{Remark}

\makeatletter
\renewcommand{\maketag@@@}[1]{\hbox{\m@th\normalsize\normalfont#1}}%
\makeatother

\begin{document}
\title{Power Allocation for Cell-Free Massive MIMO ISAC Systems with OTFS Signal}

\author{Yifei Fan,~\IEEEmembership{Graduate Student Member,~IEEE,}
		Shaochuan Wu,~\IEEEmembership{Senior Member,~IEEE,}\\
        Xixi Bi and
        Guoyu Li,~\IEEEmembership{Student Member,~IEEE}

\thanks{This work was supported by the National Natural Science Foundation of China under Grant 62271167. (\emph{Corresponding author: Shaochuan Wu.})}

\thanks{The authors are with the School of Electronics and Information Engineering, Harbin Institute of Technology, Harbin 150001, China (e-mail: yifan@stu.hit.edu.cn; scwu@hit.edu.cn; xxbi@stu.hit.edu.cn; lgy@stu.hit.edu.cn).}

}

\maketitle

\begin{abstract}

Applying integrated sensing and communication (ISAC) to a cell-free massive multiple-input multiple-output (CF mMIMO) architecture has attracted increasing attention. This approach equips CF mMIMO networks with sensing capabilities and resolves the problem of unreliable service at cell edges in conventional cellular networks. However, existing studies on CF-ISAC systems have focused on the application of traditional integrated signals. To address this limitation, this study explores the employment of the orthogonal time frequency space (OTFS) signal as a representative of innovative signals in the CF-ISAC system, and the system's overall performance is optimized and evaluated. A universal downlink spectral efficiency (SE) expression is derived regarding multi-antenna access points (APs) and optional sensing beams. To streamline the analysis and optimization of the CF-ISAC system with the OTFS signal, we introduce a lower bound on the achievable SE that is applicable to OTFS-signal-based systems. Based on this, a power allocation algorithm is proposed to maximize the minimum communication signal-to-interference-plus-noise ratio (SINR) of users while guaranteeing a specified sensing SINR value and meeting the per-AP power constraints. The results demonstrate the tightness of the proposed lower bound and the efficiency of the proposed algorithm. Finally, the superiority of using the OTFS signals is verified by a 13-fold expansion of the SE performance gap over the application of orthogonal frequency division multiplexing signals. These findings could guide the future deployment of the CF-ISAC systems, particularly in the field of millimeter waves with a large bandwidth.

\end{abstract}


\section{Introduction}
\IEEEPARstart{D}{riven} by global demands for spectrum congestion alleviation and device size miniaturization, integrated sensing and communication (ISAC) has emerged as an effective solution~\cite{liu2020joint,Fang2023Joint}. The fundamental concept of an ISAC system relies on the shared infrastructure, resources, and signals between communication and sensing (C\&S) to enhance spectral efficiency (SE) and sensing performance while reducing hardware and deployment costs~\cite{cui2021integrating}. Moreover, the ISAC also advocates a more profound integration paradigm where the two functionalities, rather than being seen as separate end-goals, are co-designed for mutual benefits~\cite{liu2022integrated}. From the network-level perspective, this co-design is expected to equip the existing cellular networks with ubiquitous perceptive capability, resulting in perceptive mobile networks~\cite{zhang2021enabling}. 

Unfortunately, due to the inherent flaws of traditional cellular networks, attached ISAC systems cannot provide reliable C\&S services for ultra-reliable applications, such as autonomous driving at cell edges~\cite{ngo2017cell,zheng2024mobile,Zhang2020Prospective}. However, cell-free massive multiple-input multiple-output (CF mMIMO), a promising technology for the future beyond-5G era, stands out as a viable solution~\cite{chen2022survey,zhang2021performance}. Due to its ability to deliver uniform, ubiquitous services, the CF mMIMO can completely eliminate the cell edges, considerably reducing the risk of service outages in autonomous driving~\cite{ammar2021user}. Further, challenges in the full-duplex system deployment have significantly motivated this technology. In the CF mMIMO architecture, geographically distributed ISAC transmitters and receivers can avoid severe self-interference in full-duplex systems where transmitting and receiving antennas are co-located; this architecture is known as multi-static sensing~\cite{wei2023integrated}. 

\begin{table*}[!t]
\centering
\caption{A Literature Overview on CF-ISAC*}
\label{tab1}
\begin{threeparttable}
\rowcolors{2}{lightgray!30}{white}
\renewcommand\arraystretch{1.5}{
    \centering
    \begin{tabular}{ccccccc}
    \Xhline{0.8pt}\rowcolor{gray!35}	
        \textbf{Ref.} & \textbf{Year} & \colorbox{gray!35}{\makecell{\textbf{Sensing}\\\textbf{period}}} & \textbf{Contributions} & \textbf{Objective function} & \colorbox{gray!35}{\makecell{\textbf{ISAC}\\\textbf{signal}}} & \makecell{\textbf{Unknown}\\\textbf{target position**}} \\ \Xhline{0.5pt}
        \cite{sakhnini2022target} & 2022 & Uplink & Protocol for uplink CF JCR & - & OFDM & ~ \\ 
        \cite{chu2023integrated} & 2023 & Downlink & {\colorbox{lightgray!30}{\makecell{ ISAC in Cell-free mMIMO \\ system with OFDM modulation}}} & - & OFDM & ~ \\ 
        \cite{cao2023joint} & 2023 & Downlink & \makecell{V-OFDM integrated signals \\ for cell-free ISAC systems} & Maximize sum-rate of UEs and targets & V-OFDM & ~ \\ 
        \cite{demirhan2023cell} & 2023 & Downlink & ISAC beamforming design & Maximize sensing SNR & SC & ~ \\ 
        \cite{behdad2024multi} & 2023 & Downlink & \makecell{Power allocation to \\ maximize the detection probability} & Maximize sensing SINR & SC & ~ \\ 
        \cite{liu2023joint} & 2023 & Downlink & {\colorbox{lightgray!30}{\makecell{BS mode selection and \\ transmit beamforming}}} & Maximize sum of sensing SINR  & SC & ~ \\ 
        \cite{zeng2023integrated} & 2023 & \makecell{Downlink\\(NAFD)} & \makecell{Deep Q-network based power \\allocation with NAFD-ISAC} & \makecell{Maximize sum communication rate \\ and minimize positioning error} & SC & ~ \\ 
        \cite{elfiatoure2023cell} & 2023 & Downlink & {\colorbox{lightgray!30}{\makecell{AP mode selection and power \\ control using the MRSR metric}}} & Maximize minimum communication SINR & SC & ~ \\ 
        \cite{mao2023communication} & 2023 & Downlink & Communication-sensing region & Maximize sum ergodic rate of UEs & SC & Y \\ 
        \cite{behdad2024joint} & 2024 & Downlink & {\colorbox{lightgray!30}{\makecell{ISAC in Cell-free mMIMO \\ systems with URLLC}}} & Minimize energy consumption & SC & ~ \\ 
        \makecell{This \\ paper} & 2024 & Downlink & \makecell{ Integrated design with OTFS \\ signal in CF-ISAC systems} & Maximize minimum communication SINR & OTFS & Y \\ \Xhline{0.8pt}
    \end{tabular}}
    \begin{tablenotes}
    \item[*] The acronyms in the table are given as JCR (joint communication and radar), mMIMO (massive MIMO), V-OFDM (vector orthogonal frequency division multiplexing), UEs (user terminals), SNR (signal-to-noise ratio), SC (single-carrier), SINR (signal-to-interference-and-noise ratio), BS (base station), NAFD (network-assisted full-duplex), MASR (mainlobe-to-average-sidelobe ratio), and URLLC (ultra-reliable low-latency communications).
    \item[**] The precise location of the target is not known throughout the precoding process.
	\end{tablenotes}
	\end{threeparttable}
\end{table*}

\subsection{Related Work}

An overview of some recent research on cell-free integrated sensing and communication (CF-ISAC) is presented in Table~\ref{tab1}, where it can be seen that \emph{single-carrier} (SC) has been mainly in use as an ISAC signal, with notable exceptions~\cite{sakhnini2022target,chu2023integrated,cao2023joint} using the orthogonal frequency division multiplexing (OFDM) signal and its variant (i.e., V-OFDM signal). Particularly, the authors in~\cite{sakhnini2022target} proposed a valuable protocol for uplink C\&S with the OFDM signal, where a set of access points (APs) were allocated for radar transmission using the same time-frequency resources as users. Further, in~\cite{chu2023integrated}, the target reflection of the communication signals was considered, and a complete signal model with the OFDM modulation was developed. However, neither~\cite{sakhnini2022target} nor~\cite{chu2023integrated} incorporated the optimization process of the overall performance. In contrast, reference~\cite{cao2023joint} introduced the V-OFDM integrated signals to the CF-ISAC systems, where the APs worked in the full-duplex mode. The maximum sum rate was used as an objective function, and a low-complexity algorithm was provided to tackle the problem of joint resource allocation. 

The aforementioned studies have been based on the application of traditional integrated signals, leaving many novel signals with great potential unexplored. This has been the main motivation for this work. This study \emph{designs} a superior ISAC signal for the CF-ISAC system and \emph{optimizes} and \emph{evaluates} overall system performance. Among various candidate signals for 6G, the orthogonal time frequency and space (OTFS) signal is extraordinary. Leveraging the sparsity of delay-Doppler (DD) domain channels, the OTFS could improve communication SE by reducing pilot overhead and showing resilience to delay and Doppler spreads. Moreover, sensing engages delay and Doppler parameters that depict the range and velocity characteristics of moving targets, which perfectly aligns with the OTFS~\cite{yuan2023new}. In addition, it has been demonstrated that the OTFS signal can achieve parameter estimation precision similar to the radar signals (e.g., frequency modulated continuous wave (FMCW)), outperforming the OFDM signals~\cite{gaudio2020effectiveness}. Particularly in high-velocity scenarios, such as autonomous driving, the OTFS is anticipated to bolster system efficiency and reliability~\cite{zheng2022cell,Wu2023OTFS}. These advantages make the OTFS an ideal signal for fully unleashing the potential of the CF-ISAC systems.

\subsection{Contributions}

Unlike prior studies, this study investigates the employment of the OTFS, a novel integrated signal, in CF-ISAC systems. For a closer view of real-world applications, the following autonomous driving circumstance is considered. Assume that the CF-ISAC network can jointly detect a certain area ahead of a vehicle as a target's sensing hotspot area~\cite{behdad2024multi}, contributing to an assessment of forthcoming road conditions. In this scenario, the assumption adopted by most prior studies (i.e., the exact location where the target can appear is known) tends to be less reasonable. Therefore, this study considers the downlink precoding process of the integrated signal while keeping the potential target's location uncertain, providing a more practical research perspective. To the best of the authors' knowledge, the integrated design and optimization of the OTFS signals in a CF-ISAC system has not been explored in the existing literature. The main contributions of this paper can be summarized as follows.
\begin{itemize}
  \item[$\bullet$] A comprehensive CF-ISAC system model with multi-antenna APs is developed based on the OTFS signal. The embedded pilot (EP)-based channel estimation and maximum-ratio (MR) precoding are used to derive a closed-form achievable SE expression for the downlink communication, considering the effects of channel estimation errors and the interference from additional sensing beams. To further analyze and optimize the SE performance of the CF-ISAC system effectively, this study proposes a low-complexity SE lower bound, which is universally applicable to all OTFS-signal-based systems.
  \item[$\bullet$] From the sensing perspective, this study inventively introduces a sensing model that operates without prior knowledge about the precise target location during the precoding process. By strategically directing optional sensing beams toward the center of the sensing hotspot area, a closed-form expression for the sensing signal-to-interference-plus-noise (SINR) is derived. 
  \item[$\bullet$] To improve the communication SINR between users while satisfying the particular sensing SINR constraint and per-AP power constraints, a max-min fairness problem is formulated. To solve this non-convex problem, this study proposes an innovative power allocation algorithm using iterative convex optimization of fractional programming. Numerical results verify the effectiveness of the proposed algorithm, which is reflected not only in a general increase in achievable SE but also in the suppression of interference components. In addition, the role of dedicated sensing beams in enhancing the system's ISAC performance is shown and explained.
  \item[$\bullet$] The results highlight the superiority of the OTFS over the OFDM signal in CF-ISAC systems from a new perspective. This study reveals a two-fold effect on the downlink SE of the OFDM-signal-based system when the number of subcarriers remains fixed while the subcarrier bandwidth increases. The numerical results indicate a significant decrease of 44.42\% in the achievable SE of the OFDM-signal-based system when the subcarrier bandwidth increases from a conventional 15\,kHz to 135\,kHz, which results in a 13-fold expansion in the SE performance gap compared to the OTFS-signal-based system.
\end{itemize}

$\emph{Notation:}$ Lowercase letters, $a$, denote scalars, boldface lowercase letters $\mathbf{a}$ and boldface uppercase letters $\mathbf{A}$ denote column vectors and matrices, respectively. The superscripts $\left( \cdot \right)^*$, $\left( \cdot \right)^T$, $\left( \cdot \right)^{-1}$, and $\left( \cdot \right)^{\dagger}$ represent the conjugate, transpose, inverse and conjugate-transpose operations, respectively. The operators $\left( \cdot \right)_N$, $\mathrm{vec}(\cdot)$, $\mathrm{Tr\left( \cdot \right)}$, and $\mathbb{E}\left\{ \cdot \right\}$ denote modulo-$N$, vectorization, trace, and expectation; $\mathbf{F}_{N}$ denotes the unitary discrete Fourier transform (DFT) matrix of size $N \times N$. The operator $\otimes$ denotes the Kronecker product of two matrices; $\left\lceil\cdot\right\rceil$ is the ceiling function which returns the smallest integer greater than or equal to the input value; The big-O notation $\mathcal{O}(\cdot)$ asymptotically describes the order of computational complexity; $\operatorname{diag}\{\cdot\}$ returns a diagonal matrix; $\operatorname{circ}\{\mathbf{x}\}$ represents a circulant matrix whose first column is $\mathbf{x}$. $\left\| \cdot  \right\|$ and $\left| \cdot \right|$ represent the vector and scalar Euclidean norms. Finally, $[\mathbf{A}]_{(i,j)}$, $[\mathbf{A}]_{(i,:)}$, $[\mathbf{A}]_{(:,j)}$ denote the $(i,j)$th entry, the $i$th row, and $j$th column of $\mathbf{A}$, respectively.

\section{System Model}

\begin{figure}[!t]
  \centering
  \includegraphics[width=3.4in]{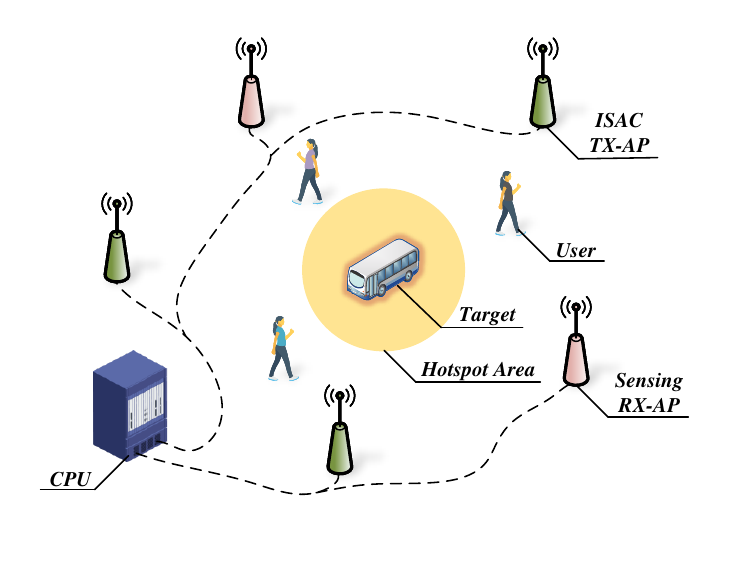}
  \caption{Illustration of the CF-ISAC system setup.}
  \label{fig:fig_1}
\end{figure}

This study considers a CF-ISAC system featuring downlink communication and multi-static sensing, as depicted in Fig.~\ref{fig:fig_1}. The OTFS is used as an integrated signal in the system. Each AP is equipped with $M_{\mathrm t}$ isotropic antennas configured in a uniform linear array (ULA). All the APs share connectivity and synchronization via fronthaul links to a centralized processing unit (CPU). In addition, each AP operates either as an ISAC transmitter or a sensing receiver. The $N_{\mathrm{tx}}$ transmitting APs jointly serve $K_{\mathrm u}$ single-antenna users by transmitting integrated OTFS signals. If desired, an extra OTFS signal can be integrated into the transmission, using the same time-frequency resources as the communication OTFS signals~\cite{behdad2024multi}. In this case, each transmitting AP steers a dedicated sensing beam toward the target's hotspot area. In contrast, $N_{\mathrm{rx}}$ APs function as sensing receivers, simultaneously receiving the potential echo signals used to detect the target.

\begin{table*}[!ht]
\begin{center}
\caption{List of Notations}
\label{tab2}
\renewcommand\arraystretch{1.5}{
\begin{tabular}{ c  c || c  c }
\Xhline{0.8pt}
\textbf{Symbol} & \textbf{Definition} & \textbf{Symbol} & \textbf{Definition}\\
\Xhline{0.6pt}
$N_{\mathrm{tx}}$ & Number of transmitting APs & 
$N_{\mathrm{rx}}$ & Number of receiving APs\\
\hline
$K_{\mathrm u}$ & Number of users &  
$M$ & Number of subcarriers\\
\hline 
$N$ & Number of symbols & 
$\Delta f$ & Subcarrier bandwidth\\
\hline 
$T$ & Symbol duration & 
$M_{\mathrm t}$ & Number of antennas per AP\\
\hline
$P_{q}^{\mathrm{dt}}$ & Uplink data transmit power at the $q$th user  & 
$P_{q}^{\mathrm{Pil}}$ & Uplink pilot transmit power at the $q$th user \\
\hline 
$\eta_q$ & Uplink power control coefficient at the $q$th user  & 
$P_{p}$ & Downlink transmit power at the $p$th transmitting AP \\
\hline 
$\eta_{pq}$ & \makecell{Downlink power control coefficient from \\ the $p$th transmitting AP to the $q$th user}  & 
$L_{pq}$ & \makecell{Number of paths from the $p$th \\ transmitting AP to the $q$th user} \\
\hline 
$\tau _{pq,i}$ & \makecell{Delay of the $i$th path from the $p$th \\ transmitting AP to the $q$th user}  & 
$\nu _{pq,i}$ & \makecell{Doppler shift of the $i$th path from \\ the $p$th transmitting AP to the $q$th user} \\
\hline 
$\tau_{\max}$ & \makecell{The maximum delay spread among all channel paths}  & 
$\nu_{\max}$ & \makecell{The maximum Doppler spread among all channel paths} \\
\hline 
$\ell _{pq,i}$ & \makecell{Delay index of the $i$th path from \\ the $p$th transmitting AP to the $q$th user} & 
$k _{pq,i}$ & \makecell{Doppler index of the $i$th path from \\ the $p$th transmitting AP to the $q$th user}\\
\hline 
$\ell _{pq}$ & \makecell{Delay tap corresponding to the largest delay \\ spread from the $p$th transmitting AP to the $q$th user} & 
$k _{pq}$ & \makecell{Doppler tap corresponding to the largest Doppler \\ spread from the $p$th transmitting AP to the $q$th user}\\
\hline 
$\kappa _{pq,i}$ & \makecell{Fractional Doppler of the $i$th path \\ from the $p$th transmitting AP to the $q$th user} & 
$X_{q}[n,m]$ & \makecell{Symbols of the $q$th user arranged over \\ the TF domain grid $\Lambda=\left\{nT,m\Delta f\right\}$}\\
\hline
$s_{q}$ & Transmitted time domain signal from the $q$th user & 
$r_{p}$ & Received time domain signal at the $p$th transmitting AP\\
\hline
$\varphi_{q}$ & The pilot symbol for the $q$th user & 
$\gamma_{\mathrm s}$ & The minimum required sensing SINR threshold \\
\hline
$\beta_{pr}$ & \makecell{Sensing channel gain from the $p$th \\ transmitting AP to the $r$th receiving AP}  & 
$\alpha_{pr}$ & \makecell{RCS of the target through the reflected path from \\ the $p$th transmitting AP to the $r$th receiving AP}\\
\hline 
$\mathbf{x}_{q}$ & Transmitted DD domain signal from the $q$th user & 
$\mathbf{y}_{p}$ & Received DD domain signal at the $p$th transmitting AP\\
\hline 
$\mathbf{a}(\varphi,\vartheta)$ & \makecell{Array response vector with the \\ azimuth angle $\varphi$ and elevation angle $\vartheta$} & 
$\mathbf{h}_{pq,i}$ & \makecell{Channel impulse response of the $i$th path \\ from the $p$th transmitting AP to the $q$th user} \\
\hline 
$\hat{\mathbf{h}}_{pq,i}$ & \makecell{The MMSE estimate of the \\ channel impulse response $\mathbf{h}_{pq,i}$} & 
$\mathbf{R}_{pq,i}$ & \makecell{Spatial correlation matrix of the $i$th path \\ from the $p$th transmitting AP to the $q$th user} \\
\hline 
$\mathbf{B}_{pq,i}$ & \makecell{The covariance matrix of \\ the MMSE estimate vector $\hat{\mathbf{h}}_{pq,i}$ } & 
$\mathbf{H}_{pq}$ & \makecell{Effective DD domain channel from \\ the $p$th transmitting AP to the $q$th user} \\
\hline 
$\mathbf{T}_{pq}^{(i)}$ & \makecell{Delay and Doppler index matrix in the DD domain of \\ the $i$th path from the $p$th transmitting AP to the $q$th user}& 
$\hat{\mathbf{H}}_{pq}$ & \makecell{The MMSE estimate of the \\ effective DD domain channel $\mathbf{H}_{pq}$} \\
\hline 
$\mathbf{H}_{pr}$ &  \makecell{Sensing desired channel of from \\ the $p$th transmitting AP to the $r$th receiving AP} & 
$\mathbf{G}_{pr}$ & \makecell{Sensing interference channel of from \\ the $p$th transmitting AP to the $r$th receiving AP}\\
\hline
$\mathbf{R}_{\mathrm{tx},(pr)}$ & \makecell{The spatial correlation matrix corresponding to the $p$th \\ transmitting AP and with respect to the $r$th receiving AP} & 
$\mathbf{R}_{\mathrm{rx},(pr)}$ & \makecell{The spatial correlation matrix corresponding to the $r$th \\ receiving AP and with respect to the $p$th transmitting AP}\\
\Xhline{0.8pt}
\end{tabular}}
\end{center}
\end{table*}

\subsection{OTFS Modulation and Communication Channel}

The OTFS signal is assumed to have $M$ subcarriers, each with a bandwidth of $\Delta f$ (Hz), and $N$ symbols with a symbol duration of $T$ (seconds). A cyclic prefix (CP) of sample length $N_{\mathrm{cp}}$ is added to the entire frame, ensuring the corresponding CP duration satisfies the condition of $T_{\mathrm{cp}}\geq\tau_{\max}$~\cite{zhang2023low}. This study focuses on extending the CF-OTFS model with single-antenna APs into the model where each AP has multiple antennas, following the work presented in~\cite{mohammadi2022cell}. The $q$th user's modulated information symbols $\begin{aligned}\{x_q[k,\ell],k=0,\ldots,N-1,\ell=0,\ldots,M-1\}\end{aligned}$ are scheduled on the DD grid $\Gamma=\left\{\frac{k}{NT},\frac{\ell}{M\Delta f}\right\}$, where $k$ and $\ell$ represent the Doppler and delay indexes, respectively. These symbols are zero-mean independent and identically distributed (i.i.d.), with the uplink transmit power of $\mathbb{E}\{|x_q[k,\ell]|^2\}=P_q^{\mathrm{dt}}$ at user $q$. After performing the inverse symplectic finite Fourier transform (ISFFT) operation, the set of $MN$ DD domain symbols $x_q[k,\ell]$ is converted to symbols $X_{q}[n,m]$ in the time-frequency (TF) domain as follows:
\begin{equation} X_{q}[n,m] = \frac {1}{\sqrt {MN}}\sum _{k=0}^{N-1}\sum _{\ell =0}^{M-1} x_{q}[k,\ell] e^{j2\pi \left ({\frac {nk}{N}-\frac {m\ell }{M}}\right)},\end{equation} 

\noindent for $n=0,\ldots,N-1$ and $m=0,\ldots,M-1$. Next, by performing the Heisenberg transform, $X_{q}[n,m]$ are converted to a time-domain signal as
\begin{align} s_q(t)=\sqrt{\eta_q}\sum_{n=0}^{N-1}{\sum_{m=0}^{M-1}{X_q}}[n,m]g_{tx}(t-nT)e^{j2\pi m\Delta f(t-nT)},\end{align}

\noindent where $0\leq\eta_q\leq1$ is the uplink power control coefficient at the $q$th user, and $g_{tx}(t)$ is the transmitting pulse-shaping filter.

This study considers the block-fading channels characterized by constant and independent channel realizations within each coherence block. Assuming that users are traveling at high velocities, the channels between the transmitting AP and users consequently experience doubly selective fading. The channel impulse response from transmitting AP $p$ to user $q$ in the DD domain is expressed as~\cite{raviteja2018interference}
\begin{equation} \mathbf{h}_{pq}(\tau,\nu) = \sum _{i=1}^{ L_{pq}} \mathbf{h}_{pq,i}\delta (\tau - \tau _{pq,i})\delta (\nu - \nu _{pq,i}),\end{equation}

\noindent where the correlated Rayleigh fading distribution is assumed, for which $\mathbf{h}_{pq,i}\sim \mathcal{CN}(\mathbf{0},\mathbf{R}_{pq,i})$ where $\mathbf{R}_{pq,i}=\mathbb{E}\{\mathbf{h}_{pq,i}^{\phantom{\dagger}}\mathbf{h}_{pq,i}^{\dagger}\}\in \mathbb{C}^{M_{\mathrm t}\times M_{\mathrm t}}$ is the spatial correlation matrix that covers the effect of geometric path loss, shadowing, and spatial correlation between antennas~\cite{bjornson2017massive}; $\tau_{pq,i}$, $\nu_{pq,i}$, and $L_{pq}$ represent the $i$th path's delay, Doppler shift, and the number of paths from transmitting AP $p$ to user $q$, respectively; the relationships between the delay and Doppler shifts and their DD domain indexes $\ell_{pq,i}=0,\ldots,M-1, k_{pq,i}=0,\ldots,N-1$ are given by $\tau_{pq,i}=\frac{\ell_{pq,i}}{M\Delta f}$ and $\nu_{pq,i}=\frac{k_{pq,i}+\kappa_{pq,i}}{NT}$, respectively, where $\kappa_{pq,i}\in(-0.5,0.5)$ denotes the fractional Doppler corresponding to the $i$th path.

\subsection{Uplink Communication Model}
During the uplink period, users send their pilot and data to the transmitting APs. The signal received at the $s$th antenna of the $p$th transmitting AP is expressed by 
\begin{align}&\hspace {-0.5pc}r_{ps}(t) =\nonumber \sum _{q=1}^{K_{\mathrm{u}}} \sqrt {\eta _{q}} \int_\tau \int_\nu h_{pq,s}(\tau,\nu)s_{q}(t-\tau)e^{j2\pi \nu (t-\tau)} d\tau d\nu \\& \qquad\qquad\qquad\qquad\qquad\qquad\qquad\qquad\qquad+\, w_{ps}(t),\end{align}

\noindent where $h_{pq,s}(\tau,\nu)$ denotes the $s$th element of $\mathbf{h}_{pq}(\tau,\nu)$ and $w_{ps}(t)\sim\mathcal{CN}(0,\sigma_n^2)$ is the noise received by the $s$th antenna of the $p$th transmitting AP in the time domain, with a noise variance $\sigma_n^2$. After performing the Wigner transform equipped with a receiving filter $g_{rx}(t)$, the received samples in the TF domain $\{Y_{ps}[n,m], n=0,\ldots,N-1, m=0,\ldots,M-1\}$ can be obtained by a sampler as follows:
\vspace{-5pt}
\begin{equation} 
Y_{ps}[n,m]=\int r_{ps}(t) g_{rx}(t-nT) e^{-j2\pi m \Delta f (t-nT)} dt.
\end{equation}

\vspace{-5pt}

Finally, after applying the SFFT to $Y_{ps}[n,m]$ and assuming non-ideal rectangular windows are used in the transmitting and receiving pulse-shaping filters, the DD domain signal received at the $s$th antenna of the $p$th transmitting AP is expressed as~\cite{raviteja2018interference}
\begin{align}
y_{ps}[k,\ell] =& \sqrt {\eta _{q}} \sum _{q=1}^{K_{\mathrm{u}}} \sum _{k'=0}^{ k_{pq}} \sum _{\ell '=0}^{ \ell _{pq}} b[k',\ell ']\!\! \sum _{c=-N/2}^{N/2-1}\!\! {h}_{pq,s}[k',\ell ']\alpha [k,l,c] \nonumber \\
&\times x_{q}[(k - k' + c)_{N},(\ell - \ell ')_{M}] + w_{ps}[k,\ell],\label{eq:y_ps}
\end{align}

\noindent where $k_{pq}$ and $\ell _{pq}$ denote the largest delay and Doppler taps between transmitting AP $p$ and user $q$, respectively. The path indicator is denoted as $b[k', \ell ']$, where $b[k', \ell ']=1$ signifies the presence of a path with Doppler tap $k'$ and delay tap $\ell '$, and $b[k', \ell ']=0$ indicates the absence of such a path. $\alpha [k,\ell,c]$ in~\eqref{eq:y_ps} is given by~\cite{raviteja2018interference}
\begin{align} \alpha [k,\ell,c] = \begin{cases} \displaystyle \frac {1}{N}\beta _{i}(c) e^{-j2\pi \frac {(\ell -\ell ')(k'+\kappa ')}{MN}},\qquad\  \ell '\leq \ell < M  \\[0.5pc] \displaystyle \frac {1}{N}(\beta _{i}(c) - 1) e^{-j2\pi \frac {(\ell -\ell ')(k'+\kappa ')}{MN}}e^{-j2\pi \frac {(k-k'+c)_{N}}{N}},\\ \hspace {5em}\qquad \qquad \qquad \qquad \ 0\leq \ell < \ell ', \end{cases}
\end{align}

\noindent where $\beta_i(c)\triangleq\frac{e^{-j2\pi(-c-\kappa^{\prime})}-1}{e^{-j\frac{2\pi}{N}(-c-\kappa^{\prime})}-1}$, and $\kappa^\prime$ denotes the corresponding fractional Doppler. In addition, $w_{ps}[k,\ell]$ in~\eqref{eq:y_ps} is the DD domain received noise, which is given by~\cite{mohammadi2022cell}
\begin{equation} w_{ps}[k,\ell]=\frac{1}{\sqrt{MN}}\sum_{n=0}^{N-1}\sum_{m=0}^{M-1}W_{ps}[n,m]e^{-j2\pi(\frac{nk}{N}-\frac{m\ell}{M})}, \end{equation} 

\noindent where $W_{ps}[n,m]$ is the sampled noise in the TF domain. It can be easily demonstrated that $w_{ps}[k,\ell]$ have the same distribution as $w_{ps}(t)$, i.e., $w_{ps}[k,\ell] \sim \mathcal{CN}(0,\sigma_n^2)$. 

Further, to simplify the expression, this study reformulates the DD domain input-output relationship into a vector form. The specific process is as follows. First, invoke the matrix form relationship in~\eqref{eq:y_ps} and rearrange them as $\mathbf{X}_q\in\mathbb{C}^{M\times N}$ and $\mathbf{Y}_p=\left[{{\mathrm{vec}(\mathbf{Y}}_{p1}),\ldots,{\mathrm{vec}(\mathbf{Y}}_{ps}),\ldots,\mathrm{vec}({\mathbf{Y}}_{pM_{\mathrm t}}})\right]\in\mathbb{C}^{MN\times M_{\mathrm t}}$, where the $(k,\ell)$th entries of matrices $\mathbf{X}_q$ and $\mathbf{Y}_{ps}$ are $x_q[k,\ell]$ and $y_{ps}[k,\ell]$, respectively. Then, the vector of transmitted symbols from user $q$ can be obtained as $\mathbf{x}_{q}=\mathrm{vec}(\mathbf{X}_q)\in \mathbb{C}^{MN\times1}$. Similarly, $\mathbf{y}_{p}=\mathrm{vec}(\mathbf{Y}_p) \in \mathbb{C}^{M_{\mathrm t}MN\times1}$ and $\mathbf{w}_{p} \in \mathbb{C}^{M_{\mathrm t}MN\times1}$ denote the received signal vector at transmitting AP $p$ and the corresponding noise vector, respectively. Consequently, the vector form of the input-output relationship can be obtained by\footnote{Note that the uplink channel is represented by $\mathbf{H}_{pq}^{\dagger}$, although there will only be a transpose operation and no additional conjugation involved in practice~\cite{bjornson2017massive}. Furthermore, since uplink is discussed less in this paper, the original notation $\mathbf{H}_{pq}$ is reserved for the downlink model.}\cite{mohammadi2022cell}
\begin{equation} {\mathbf{y}}_{p} = \sum _{q=1}^{K_{\mathrm{u}}} \sqrt { \rho _{q}^{\mathrm {dt}}\eta _{q}} {\mathbf{H}}_{pq}^{\dagger} \tilde {\mathbf{x}}_{q} + {\mathbf{w}}_{p},\end{equation}

\noindent where $\tilde{\mathbf{x}}_{q}=\frac{1}{\sqrt{P_{q}^{\mathrm{dt}}}}\mathbf{x}_{q}\in\mathbb{C}^{MN\times1}$; $\rho_{q}^{\mathrm{dt}}=\frac{P_{q}^{\mathrm{dt}}}{\sigma_{n}^{2}}$ is the normalized uplink signal-to-noise ratio (SNR). Let ${\mathbf{H}}_{pq,s}\in \mathbb{C}^{MN\times MN}$ be an effective DD domain channel between the $q$th user and the $s$th antenna of the $p$th transmitting AP; then, the concatenated channel ${\mathbf{H}}_{pq}=\left[{{\mathbf{H}}_{pq,1},\ldots,{\mathbf{H}}_{pq,s},\ldots,{\mathbf{H}}_{pq,M_{\mathrm t}}}\right] \in \mathbb{C}^{MN\times M_{\mathrm t}MN}$ is given by~\cite{li2021performance}
\begin{equation} {\mathbf{H}}_{pq} = \sum _{i=1}^{ L_{pq}} \left( \mathbf h_{pq,i}^T\otimes {\mathbf{T}} _{pq}^{(i)} \right),\label{eq:DD_channel}\end{equation}

\noindent where $\mathbf{T}_{pq}^{(i)}=(\mathbf{F}_N\otimes\mathbf{I}_M)\mathbf{\Pi}^{\ell_{pq,i}}\boldsymbol{\Delta}^{k_{pq,i}+\kappa_{pq,i}}(\mathbf{F}_N^\dagger\otimes\mathbf{I}_M)$, in which $\boldsymbol{\Pi}=\text{circ}\{[0,1,0,\ldots,0]_{MN\times1}^T\}$ denotes an $MN\times MN$ permutation matrix and $\boldsymbol{\Delta}=\operatorname{diag}\{z^0,z^1,\ldots,z^{MN-1}\}$, with $z=e^{\frac{j2\pi}{MN}}$. 

The symbol transmitted from the $q$th user, denoted by $\tilde{\mathbf{x}}_{q}$, is detected by the $p$th transmitting AP through multiplication of the received signal $\mathbf{y}_{p}$ with the Hermitian transpose of the locally obtained channel estimation matrix, $\hat{\mathbf{H}}_{pq}$. Please refer to Section~\ref{sec:Channel Estimation} for a detailed explanation of the channel estimation process.

\subsection{Downlink Communication Model}

In the downlink process, the transmitting APs employ MR precoding to transmit integrated signals to serve $K_{\mathrm{u}}$ users. Assume $\mathbf{s}_{q}=\mathrm{vec}(\mathbf{S}_{q})\in\mathbb{C}^{MN\times1}$ is a zero-mean downlink communication signal vector for user $q$, where $\mathbf{S}_q\in\mathbb{C}^{M\times N}$ represent the DD domain modulated symbols, whose $(k,\ell)$th entry $s_{q}[k,\ell]$ represents the symbol on the $k$th Doppler and $\ell$th delay grid. For the sake of notation simplicity, the optional dedicated sensing signal is expressed in the same form as $\mathbf{s}_{0}$. 

This study assumes that the sensing signal and users' data signals are independent of each other to simplify the analysis~\cite{behdad2024multi}. Therefore, the signal transmitted from transmitting AP $p$ can be expressed by
\begin{equation} {\mathbf{x}}_{p}= \sqrt {\rho_{\mathrm{d}}} \sum _{q=0}^{K_{\mathrm{u}}} \eta _{pq}^{1/2}\hat {\mathbf{H}}_{pq}^{\dagger} {\mathbf{s}} _{q},\label{eq:x_p}\end{equation}

\noindent where $\rho_{\mathrm{d}}=\frac{P_\mathrm{d}}{\sigma_{n}^{2}}$ is the normalized downlink SNR; $\eta _{pq},\, p=1,\ldots,N_{\mathrm{tx}},\,q=0,\ldots,K_{\mathrm{u}}$ are the power control coefficients set to make the average transmit power $\rho_p$ at each transmitting AP satisfy the following power constraint~\cite{ngo2017cell}
\begin{equation} \rho_p=\frac{\mathbb {E}\left \{{\| {\mathbf{x}}_{p}\|^{2}}\right \}}{MN} \leq \rho_{\mathrm{d}}.\label{eq:power_constraint}\end{equation}

Based on the effective channel matrix, the signal received by user $q$ in the DD domain is modeled as
\begin{align} {\mathbf{z}}_{q}=&\sum _{p=1}^{N_{\mathrm{tx}}}{\mathbf{H}}_{pq} {\mathbf{x}}_{p} + {\mathbf{w}}_{q} \nonumber \\[-4pt]
=&\sqrt {\rho_{\mathrm{d}}}\sum _{p=1}^{N_{\mathrm{tx}}} \eta _{pq}^{1/2} {\mathbf{H}}_{pq} \hat {\mathbf{H}}_{pq}^{\dagger} {\mathbf{s}} _{q} \nonumber \\[-4pt]
&+\, \sqrt {\rho_{\mathrm{d}}}\sum _{p=1}^{N_{\mathrm{tx}}} \sum _{q'\neq q}^{K_{\mathrm{u}}} \eta _{pq'}^{1/2}{\mathbf{H}}_{pq}\hat {\mathbf{H}}_{pq'}^{\dagger} {\mathbf{s}} _{q'} + {\mathbf{w}}_{q},\label{eq:z_q}\end{align}

\noindent where $\mathbf{w}_{q}\in\mathbb{C}^{MN\times1}$ is the noise vector at the user $q$.

\subsection{Sensing Model}

Consider the multi-static sensing model presented in~\cite{behdad2024multi} and assume half-wavelength-spaced antennas on each AP. Then, the antenna array response for an azimuth angle $\varphi$ and an elevation angle $\vartheta$, denoted by $\mathbf{a}(\varphi,\vartheta)\in\mathbb{C}^{M_{\mathrm t}\times1}$, is given as~\cite{bjornson2017massive}
\begingroup\makeatletter\def\f@size{9}\check@mathfonts
\begin{equation} \mathbf{a}(\varphi,\vartheta)=\frac{1}{\sqrt{M_{\mathrm t}}}\left[ 1, e^{j\pi\sin(\varphi)\cos(\vartheta)}, \ldots, e^{j(M_{\mathrm t}-1)\pi\sin(\varphi)\cos(\vartheta)}\right]^T.\end{equation}
\endgroup

\noindent Further, using $\mathbf{a}(\varphi,\vartheta)$, the array response vectors from the $p$th transmitting AP to the target and from the target to the $r$th receiving AP can be expressed by $\mathbf{h}_{pt}=\mathbf{a}(\varphi_p,\vartheta_p)$ and $\mathbf{h}_{tr}=\mathbf{a}(\phi_r,\theta_r)$, respectively. Here, $\varphi_p$ and $\vartheta_p$ are the azimuth and elevation angles from a transmitting AP $p$ to the target location, respectively. Similarly, $\phi_r$ and $\theta_r$ are the azimuth and elevation angles from the target location to receiving AP~$r$, respectively.

Considering the uncertainty of the target location in the precoding process and pointing the dedicated sensing beam to the center of the hotspot area with azimuth angle ${\varphi}_p^{\prime}$ and elevation angle ${\vartheta}_p^{\prime}$ yield the conjugate beamforming, which is expressed as follows~\cite{demirhan2023cell}:
\begin{equation} 
\hat{\mathbf{h}}_{pt}=\mathbf{a}({\varphi}_p^{\prime},{\vartheta}_p^{\prime}),\label{eq:sensing_BF}
\end{equation}

\noindent which aims at obtaining the maximum sensing power.\footnote{The strategy known as nullspace beamforming~\cite{buzzi2019using} can also be employed. However, when a local precoding technique such as MR precoding is used, the number of spatial degrees of freedom it can offer equals the number of antennas on a single transmitting AP $M_{\mathrm t}$, which is typically smaller than the number of users $K_{\mathrm{u}}$. Consequently, satisfying the conditions for generating the nullspace becomes challenging, causing the nullspace beamforming to deteriorate into the conjugate beamforming.}

When the target exists, both desired and undesired signals are received at each receiving AP. The former denotes the echo signals from the target, and the latter is independent of the target's presence. For simplicity, this study assumes the desired signals comprise line-of-sight (LOS) links between the target and each transmitting or receiving AP, while the non-line-of-sight (NLOS) links are neglected~\cite{behdad2024multi}. This study adopts the assumption that the transmitted ISAC signal, denoted by $\mathbf{s}_{q}$, is available at the CPU~\cite{zeng2023integrated}. This presumption indicates that the communication symbols also contribute to sensing via the reflected paths toward the target, so they are regarded as desired signals within the context of target sensing.

The sensing undesired signals include the thermal noise and NLOS clutter caused by temporary obstacles, while the NLOS paths caused by the permanent obstacles and LOS paths between transmitting and receiving APs are canceled out and removed in signal processing~\cite{behdad2024multi}. Let $\mathbf{G}_{pr}\in\mathbb{C}^{M_\mathrm{t}MN\times M_\mathrm{t}MN}$ denotes an unknown NLOS channel matrix between the $p$th transmitting AP and the $r$th receiving AP, which represents the reflected paths through temporary obstacles and is referred to as a target-free channel. Upon the canceling of known undesired parts, the received signal $\mathbf{y}_r\in\mathbb{C}^{M_\mathrm{t}MN\times1}$ at a receiving AP $r$ in the target's presence is formulated as
\begin{align}
\mathbf{y}_r=\sqrt{\rho_{\mathrm{d}}}\sum_{p=1}^{N_{\mathrm{tx}}}\sum_{q=0}^{K_{\mathrm{u}}}\!\Big(\underbrace{\eta _{pq}^{1/2}{\mathbf{H}}_{pr}^{\phantom{\dagger}} \hat {\mathbf{H}}_{pq}^{\dagger}\mathbf{s}_{q}}_{\text{Sensing desired part}}+\!\!\underbrace{\eta _{pq}^{1/2}{\mathbf{G}}_{pr}^{\phantom{\dagger}} \hat {\mathbf{H}}_{pq}^{\dagger}\mathbf{s}_{q}}_{\text{Sensing interference part}}\!\!\!\Big)\!+\mathbf{w}_r,
\label{eq:sensing_received}
\end{align}

\noindent where $\mathbf{H}_{pr}\triangleq\alpha_{pr}\beta^{1/2}_{pr}\big(\mathbf{h}_{tr}^{\phantom{T}}\mathbf{h}_{pt}^{T}\otimes \mathbf{T}_{pr}\big)$ in the first term is the channel matrix that depicts the reflected path through the target. Here, $\alpha_{pr}$ is the unknown radar cross-section (RCS) of the target via the reflected path from the $p$th transmitting AP to the $r$th receiving AP. The RCS is modulated by the Swerling-I model, where $\alpha_{pr}\sim\mathcal{CN}(0,\sigma_{pr}^{2})$ remains constant during the sensing period. Define $d_{pt}$ and $d_{tr}$ as the distances from the $p$th transmitting AP and the $r$th receiving AP to the target location, respectively. Then, by using the radar range equation derived in~\cite[Chap. 2]{richards2010principles}, the channel gain of the reflected path from the $p$th transmitting AP via the target to the $r$th receiving AP, denoted by $\beta_{pr}$, can be computed by
\begin{equation}
\beta_{pr}=\frac{\lambda_{c}^{2}G_{\mathrm{t}}G_{\mathrm{r}}}{(4\pi)^{3}d_{pt}^{2}d_{tr}^{2}},
\end{equation}

\noindent where $\lambda_{c}$ is the carrier wavelength; $G_{\mathrm{t}}$ and $G_{\mathrm{r}}$ denote the antenna gains at the transmitting and receiving APs, respectively. The last term $\mathbf{w}_r\sim\mathcal{CN}(\mathbf{0},\mathbf{I}_{M_\mathrm{t}MN})$ in \eqref{eq:sensing_received} is the normalized receiver noise at the $M_\mathrm{t}$ antennas of the $r$th receiving AP.

In adherence to~\cite{behdad2024multi}, the channel $\mathbf{G}_{pr}$ is modeled using the correlated Rayleigh fading model represented by
\begin{equation}
\mathbf{G}_{pr}=\bigg[\mathbf{R}_{\mathrm{rx},(pr)}^{\frac12}\mathbf{W}_{\mathrm{ch},(pr)}\left(\mathbf{R}_{\mathrm{tx},(pr)}^{\frac12}\right)^T\bigg]\otimes\mathbf{T}_{pr},
\end{equation}

\noindent where matrix $\mathbf{W}_{\mathrm{ch},(pr)}\in\mathbb{C}^{M_\mathrm{t}\times M_\mathrm{t}}$ is with i.i.d. random entries with the $\mathcal{CN}(0,1)$ distribution; matrix $\mathbf{R}_{\mathrm{rx},(pr)}\in\mathbb{C}^{M_\mathrm{t}\times M_\mathrm{t}}$ is the spatial correlation matrix corresponding to the $r$th receiving AP with respect to the direction of the $p$th transmitting AP; similarly, $\mathbf{R}_{\mathrm{tx},(pr)}\in\mathbb{C}^{M_\mathrm{t}\times M_\mathrm{t}}$ is the spatial correlation matrix for the $p$th transmitting AP in relation to the direction of the $r$th receiving AP; the spatial correlation matrix includes the channel gain, which is determined by the geometric path loss and shadowing.

After collecting the received signals $\mathbf{y}_r$ forwarded by each receiving AP $r$, for $r=1,\ldots,N_{\mathrm{rx}}$, the overall sensing signal can be concatenated into $\mathbf{y}=\left[\mathbf{y}_{1}^{T},\cdots,\mathbf{y}_{N_{\mathrm{rx}}}^{T}\right]^{T}\in\mathbb{C}^{N_{\mathrm{rx}}M_{\mathrm{t}}MN\times1}$ at the CPU.

\section{Channel Estimation}
\label{sec:Channel Estimation}

In this study, the channel estimation process is performed using an EP-based method, where a guard region comprised of null symbols is situated around each user's pilot symbol. To reduce the overhead of the guard region, users are authorized to transmit data symbols over the pilot and guard DD grids belonging to the other users~\cite{mohammadi2022cell}. Define $\ell_{\max}=\max_{p,q}\ell_{pq}$ and $k_{\max}=\max_{p,q}k_{pq}$ as the maximum delay and Doppler taps between the channel paths. Next, let $\varphi_q[k_q,\ell_q]$, with $\mathbb{E}\{|\varphi_q[k_q,\ell_q]|^2\}=P_q^{\mathrm{Pil}}$, denote the $q$th user's known pilot symbol at a specific DD grid~$[k_q,\ell_q]$, and $x_{dq}[k,\ell]$, with $\mathbb{E}\{|x_{dq}[k,\ell]|^2\}=P_q^{\mathrm{dt}}$, denote the $q$th user's data symbol at DD grid $[k,\ell]$. Then, the pilot, guard, and data symbols of the $q$th user in the DD grid are configured as follows~\cite{raviteja2019embedded}:
\begin{align} x_{q}[k,\ell] = \begin{cases} \varphi _{q} & k=k_{q}, ~\ell =\ell _{q}, \\ 0 & k\in \mathcal {K}, ~k\neq k_{q}~\ell \in \mathcal {L}, ~\ell \neq \ell _{q},\\ x_{dq}[k,\ell] & \text {otherwise}, \end{cases}\end{align}

\noindent where $\mathcal{K}=\{k_q-2k_{\max}-2\hat{k},\cdots,k_q+2k_{\max}+2\hat{k}\},\,\mathcal{L}=\{\ell_{q}-\ell_{\max}\leq\ell\leq\ell_{q}+\ell_{\max}\}$, in which $\hat{k}$ denotes the extra guard introduced to alleviate the spread of fractional Doppler. At the transmitting AP side, the received symbols $\mathbf{y}_p[k,\ell]$, for $k_q-k_{\max}-\hat{k}<k<k_q+k_{\max}+\hat{k}$, $\ell_{q}\leq\ell\leq\ell_{q}+\ell_{\max}$, are collected for channel estimation. Accordingly, reformulating the antenna-wise received pilot signal in \eqref{eq:y_ps} as $\mathbf{y}_p[k,\ell]=[y_{p1}[k,\ell], \ldots, y_{pM_\mathrm{t}}[k,\ell]]^{T}\in\mathbb{C}^{M_\mathrm{t}\times 1}$ yields~\cite{mohammadi2022cell}

\begin{equation}\begin{aligned}&\hspace {-0.5pc}\mathbf{y}_{p}[k,\ell] = \sqrt {\eta _{q}}\varphi _{q} \tilde {b}[\ell -\ell _{q}] \tilde{\mathbf{h}}_{pq}[(k-k_{q})_{N},(\ell -\ell _{q})_{M}] \\&\qquad\qquad\qquad\qquad+\,\boldsymbol{\mathcal{I}}_{1}(k,\ell)+\boldsymbol{\mathcal{I}}_{2}(k,\ell)+ \mathbf{w}_{p}[k,\ell],\label{eq:received_pilot}\end{aligned}\end{equation}

\noindent where
\begin{align} \tilde {b}[\ell -\ell _{q}] = \begin{cases} 1, & \displaystyle \sum \nolimits _{k'=0}^{k_{pq}} b[k',\ell -\ell _{q}]\geq 1 \\ 0, & \text {otherwise}, \end{cases}\end{align}

\noindent denotes the indicator of a path, and
\begin{align}&\hspace {-0.5pc}\tilde{\mathbf{h}}_{pq}[(k-k_{q})_{N},(\ell - \ell _{q})_{M}] \nonumber\\&=\sum _{k'=0}^{k_{pq}} b[k',\ell - \ell _{q}]\alpha [k,\ell,(k_q+k^{\prime}-k)_N] \mathbf{h}_{pq}[k',\ell - \ell _{q}].\end{align}

In \eqref{eq:received_pilot}, $\boldsymbol{\mathcal{I}}_{1}(k,\ell)$ represents the interference spread from the $q$th user's data symbols caused by the fractional Doppler, which can be expressed by
\begin{align}&\boldsymbol{\mathcal{I}}_{1}(k,\ell) = \sqrt {\eta _{q}}\sum _{k'=0}^{k_{pq}} \sum _{\ell '=0}^{\ell _{pq}} b[k',\ell '] \sum _{c\not \in \mathcal {K}} \mathbf{h}_{pq}[(k \!- \!k')_{N},(\ell \!- \!\ell ')_{M}] \nonumber
\\&\qquad\qquad\times \, \alpha (k,\ell,c)x_{dq}[(k - k' + c)_{N},(\ell - \ell ')_{M}].\label{eq:estimation_I1}\end{align}

\noindent Meanwhile, $\boldsymbol{\mathcal{I}}_{2}(k,\ell)$ represents the interference spread from other users' data symbols, and it is given by
\begin{align}&\hspace {-0.5pc}\boldsymbol{\mathcal{I}}_{2}(k,\ell) = \sum _{q'\neq q}^{K_{\mathrm{u}}}\sqrt {\eta _{q^{\prime} }} \sum _{k'=0}^{k_{pq'}} \sum _{\ell '=0}^{\ell _{pq'}} b[k',\ell '] \sum _{c=-N/2}^{N/2} \mathbf{h}_{pq'}[k',\ell '] \nonumber\\&\qquad\qquad\times \, \alpha (k,\ell,c)x_{dq'}[(k-k'+c)_{N},(\ell -\ell ')_{M}].\label{eq:estimation_I2}\end{align}

In spite of the across-frame-variations in the channel gain $\mathbf{h}_{pq,i}$, parameters $\ell_{pq,i}$ and $k_{pq,i}$ maintain constant over multiple OTFS frames~\cite{mohammadi2022cell}. Therefore, by leveraging the structure of the received pilot signals~\eqref{eq:received_pilot}, channel parameters can be efficiently estimated by the minimum mean-square error (MMSE) estimation.

\begin{lemma}\label{lem1}
The MMSE estimate of a channel vector $\mathbf{h}_{pq,i}$ and its variance matrix via the EP-based channel estimation are respectively given by
\begin{subequations}
\begin{align}\hat{\mathbf{h}}_{pq,i}&=\sqrt{P_{q}^{\mathrm{Pil}}\eta_{q}}\mathbf{R}_{pq,i}\mathbf{\Psi}_{pq,i}^{-1}\mathbf{y}_p\left[ k,\ell \right],\\
\mathbf{B}_{pq,i} &\triangleq \mathbb{E} \{\hat{\mathbf{h}}_{pq,i}^{\phantom{\dagger}}\hat{\mathbf{h}}_{pq,i}^{\dagger}\}=P_{q}^{\mathrm{Pil}}\eta _q\mathbf{R}_{pq,i}\mathbf{\Psi}_{pq,i}^{-1}\mathbf{R}_{pq,i},\end{align}\label{eq:estimation_final}%
\end{subequations}

\begin{figure*}[hb] 
\xhrulefill {black}{1pt}
  \centering
    \begin{small}
    \begin{equation}
    \begin{aligned}
        \mathbf{\Psi}_{pq,i}= P_{q}^{\mathrm{Pil}}\eta_q\mathbf{R}_{pq,i}+\frac1N\sum_{q^{\prime}=1}^{K_{\mathrm{u}}}\eta_q^{\prime}P_{q^{\prime}}^\mathrm{dt}\sum_{i=1}^{L_{pq^{\prime}}}\mathbf{R}_{pq^{\prime},i}-P_q^\mathrm{dt}\frac{(4k_{\max}+4\hat{k}+1)}{N^2}\sum_{i=1}^{L_{pq}}\mathbf{R}_{pq^{\prime},i}+\sigma _{n}^{2}\mathbf{I}_{N_{\mathrm{tx}}},
    \label{eq:estimated_psi}
    \end{aligned}
    \end{equation}
    \end{small}

\xhrulefill {black}{1pt}
  \centering
  \begin{small}
  \begin{equation}
  \begin{aligned}
      z_{qv}=&\sqrt{\rho_{\mathrm{d}}}\mathbb{E} \underbrace{\left\{ \sum_{p=1}^{N_{\mathrm{tx}}}{\eta _{pq}^{1/2}}[\mathbf{H}_{pq}]_{(v,:)}[\hat{\mathbf{H}}_{pq}^{\dagger}]_{(:,v)} \right\} }_{\text{Desired signal}\triangleq \mathbb{DS}_{qv}}s_{qv}+\underbrace{\sqrt{\rho_{\mathrm{d}}}\left( \sum_{p=1}^{N_{\mathrm{tx}}}{\eta _{pq}^{1/2}}[\mathbf{H}_{pq}]_{(v,:)}[\hat{\mathbf{H}}_{pq}^{\dagger}]_{(:,v)}-\mathbb{E} \left\{ \sum_{p=1}^{N_{\mathrm{tx}}}{\eta _{pq}^{1/2}}[\mathbf{H}_{pq}]_{(v,:)}[\hat{\mathbf{H}}_{pq}^{\dagger}]_{(:,v)} \right\} \right) s_{qv}}_{\text{Precoding gain uncertainty}\triangleq \mathbb{BU} _{qv}}
      \\
      &+\underbrace{\sqrt{\rho_{\mathrm{d}}}\sum_{p=1}^{N_{\mathrm{tx}}}{\sum_{v^{\prime}\ne v}^{MN}{\eta _{pq}^{1/2}}}\left[ \mathbf{H}_{pq} \right] _{(v,:)}[\hat{\mathbf{H}}_{pq}^{\dagger}]_{\left( :,v^{\prime} \right)}s_{qv^{\prime}}}_{\text{Inter-symbol interference} \triangleq \mathbb{I} _{qv1}}+\underbrace{\sqrt{\rho_{\mathrm{d}}}\sum_{\begin{subarray}{c}q^{\prime}\ne q\\q^{\prime}=0\\\end{subarray}}^{K_{\mathrm{u}}}{\sum_{p=1}^{N_{\mathrm{tx}}}{\sum_{v^{\prime}=1}^{MN}{\eta _{pq^{\prime}}^{1/2}}}}\left[ \mathbf{H}_{pq} \right] _{(v,:)}[\hat{\mathbf{H}}_{pq^{\prime}}^{\dagger}]_{\left( :,v^{\prime} \right)}s_{q^{\prime}v^{\prime}}}_{\begin{subarray}{l}\qquad\ \text{Inter-user interference}\\ \text{(including sensing interference)}\\ \end{subarray} \triangleq \mathbb{I} _{qv2}}+\underbrace{w_{qv}}_{\text{Noise}},
    \label{eq:received_DL_signal}
  \end{aligned}
  \end{equation}
  \end{small}

\xhrulefill {black}{1pt}
  \centering
  \begin{small}
  \begin{equation}
    R_{q}=\frac{\omega_{\mathsf{otfs}}}{MN}\sum_{v=1}^{MN}\log_2\left( 1+\frac{\rho_{\mathrm{d}}\left( \sum\limits_{p=1}^{N_{\mathrm{tx}}}{\sum\limits_{i=1}^{L_{pq}}{\eta _{pq}^{1/2}}}\mathrm{Tr}\left( \mathbf{B}_{pq,i} \right) \right) ^2}{\rho_{\mathrm{d}}\left( \sum\limits_{p=1}^{N_{\mathrm{tx}}}{\sum\limits_{i=1}^{L_{pq}}{\eta _{pq}}\left( \sum\limits_{j=1}^{L_{pq}}{\left( \chi _{pq}^{ij}+\kappa _{pq}^{ij} \right) \mathrm{Tr}\left( \mathbf{R}_{pq,i}\mathbf{B}_{pq,j} \right)}+\sum\limits_{q^{\prime}\ne q, q^{\prime}=0}^{K_{\mathrm{u}}}{\sum\limits_{j=1}^{L_{pq^{\prime}}}{\frac{\eta _{pq^{\prime}}}{\eta _{pq}}}}\mathrm{Tr}\left( \mathbf{R}_{pq,i}\mathbf{B}_{pq^{\prime},j} \right) \right)} \right) +1} \right) ,
  \label{eq:DL_SE_old}
  \end{equation}
  \end{small}
\end{figure*}

\vspace{-10pt}

\noindent where $\mathbf{\Psi}_{pq,i}$ is defined in \eqref{eq:estimated_psi}, which is shown at the bottom of the page.
\end{lemma}

\begin{IEEEproof}
  The proof is given in Appendix~\ref{app:lem1}.
\end{IEEEproof} 

Specifically, by recalling~\eqref{eq:sensing_BF}, the beamforming matrix for the dedicated sensing beam can be defined as $\mathbf{B}_{p0}\triangleq\hat{\mathbf{h}}_{pt}^{\phantom{\dagger}}\hat{\mathbf{h}}_{pt}^{\dagger}=\mathbf{a}(\varphi_p^{\prime},\vartheta_p^{\prime})\mathbf{a}(\varphi_p^{\prime},\vartheta_p^{\prime})^{\dagger}$.

\section{Communication Performance}

This section derives the closed-form expression for the downlink multi-antenna SE considering imperfect channel state information (CSI) and establishes a low-complexity lower bound for this expression. To this end, it is assumed that the low-complexity DD domain detector (LCD) proposed in~\cite{pandey2021low} is used for symbol detection at users, which is characterized by separate detection for each information symbol.

\subsection{Downlink SE Analysis}

This study follows the assumption that each user has access to the channel statistics but not to the channel realizations~\cite{ngo2017cell}. Next, let $v = kM + \ell$, and the i.i.d. information symbols intended for the $q$th user in the DD domain be rearranged as $s_{qv} = s_q[k, \ell]$, for $k=0,\ldots,N-1,\, \ell=0,\ldots,M-1$, which satisfies the condition of $s_{qv}\sim\mathcal{CN}(0,1)$. Accordingly, the signal received at the $q$th user \eqref{eq:z_q} can also be rearranged as \eqref{eq:received_DL_signal}, as shown at the bottom of the page. By treating the first term in \eqref{eq:received_DL_signal} as a desired signal and the other terms as uncorrelated Gaussian noise, the key result on the downlink achievable SE is obtained as follows.

\begin{theorem}\label{thm1}
For an OTFS-signal-based CF-ISAC system, an achievable downlink SE of the $q$th user is given by \eqref{eq:DL_SE_old}, as shown at the bottom of the page, where $\chi_{pq}^{ij}=\left|[\mathbf{T}_{pq}^{(i)}\mathbf{T}_{pq}^{(j)^{\dagger}}]_{(v,v)}\right|^2;$ $\kappa _{pq}^{ij}=\left| \sum\nolimits_{v^{\prime}\ne v}^{MN}{\bigl[ \mathbf{T}_{pq}^{(i)}\mathbf{T}_{pq}^{(j)^{\dagger}} \bigr] _{(v,v^{\prime})}} \right|^2$ and the pre-log factor is $\omega_{\mathsf{otfs}}=\left(1-\frac{N_{\mathrm{cp}}}{MN}\right)$.
\end{theorem}
\begin{IEEEproof}
  The proof is given in Appendix~\ref{app:thm1}.
\end{IEEEproof} 

Although Theorem~\ref{thm1} presents a closed-form expression for the calculation of downlink SE, it is critical to note that the computational complexity of this expression scales as $\mathcal{O}\big(M^{3}N^{3}\sum_{p=1}^{N_{\mathrm{tx}}}\sum_{q=0}^{K_{\mathrm{u}}}L_{pq}^2\big)$ for all $K_{\mathrm{u}}$ users. Namely, this complexity, exacerbated by the high dimensionality of matrix $\mathbf{T}_{pq}^{(i)}$ in practical implementation, could hinder effective SE analysis and optimization. Therefore, there is an urgent need for an alternative expression with reduced complexity.\footnote{In the downlink transmission, there are three distinct concepts of complexity: precoding complexity, detector complexity, and optimization complexity. Although the first two complexities have been studied in~\cite{pandey2021low}, the optimization complexity has not been comprehensively explored. This study aims to address this high-complexity problem, thereby streamlining the calculation, evaluation, and optimization of the SE performance.}

\begin{corollary}\label{cor1}
  By setting $\chi_{pq}^{ij}+\kappa_{pq}^{ij}=1$, a lower bound of the achievable downlink SE in \eqref{eq:DL_SE_old} can be obtained by \eqref{eq:DL_SE}, which is shown at the bottom of the page.\footnote{Please note that a similar SE lower bound for the minimum mean-squared error-based successive interference cancellation (MMSE-SIC) detector was proposed in~\cite{mohammadi2023cell}. However, due to the high complexity of the MMSE-SIC detector, further study of the SE lower bound for the LCD is still needed.}
\end{corollary}

\begin{IEEEproof}
  The proof is given in Appendix~\ref{app:cor1}.
\end{IEEEproof} 

\begin{remark}
  The setting $\chi_{pq}^{ij}+\kappa_{pq}^{ij}=1$ in Corollary~\ref{cor1} is fulfilled when the paths between transmitting AP $p$ and user $q$ have different delay indices, that is, $\ell_{pq,i}\neq\ell_{pq,j}$. This condition can be readily achieved if the number of OTFS subcarriers $M$ is large enough. Consequently,~\eqref{eq:DL_SE} actually defines a tight lower bound, which will be corroborated by the numerical results in Section~\ref{sec:Numerical Results}.
\end{remark}

\begin{figure*}[hb] 
\xhrulefill {black}{1pt}
  \centering
  \begin{small}
  \begin{equation}
    R_{q}=\omega _{\mathsf{otfs}}\log _2\left( 1+\underbrace{\frac{\rho_{\mathrm{d}}\left( \sum_{p=1}^{N_{\mathrm{tx}}}{\sum_{i=1}^{L_{pq}}{\eta _{pq}^{1/2}}}\mathrm{Tr}\left( \mathbf{B}_{pq,i} \right) \right) ^2}{\rho_{\mathrm{d}}\sum_{p=1}^{N_{\mathrm{tx}}}{\sum_{i=1}^{L_{pq}}{\eta _{pq}}\left( \sum_{j=1}^{L_{pq}}{\mathrm{Tr}\left( \mathbf{R}_{pq,i}\mathbf{B}_{pq,j} \right)}+\sum_{q^{\prime}\ne q, q^{\prime}=0}^{K_{\mathrm{u}}}{\sum_{j=1}^{L_{pq^{\prime}}}{\frac{\eta _{pq^{\prime}}}{\eta _{pq}}}}\mathrm{Tr}\left( \mathbf{R}_{pq,i}\mathbf{B}_{pq^{\prime},j} \right) \right)}+1}}_{\displaystyle{\triangleq \text{SINR}^{(\mathrm c)}_q}} \right),
  \label{eq:DL_SE}
  \end{equation}
  \end{small}
\end{figure*}

\subsection{OFDM-Signal-Based System}

The numerical results presented in this study compare the performance of the OTFS-signal-based system with its OFDM counterpart. As elucidated in~\cite{raviteja2018interference}, the OTFS employs the Heisenberg transform instead of the inverse fast Fourier transform (IFFT) used in the OFDM signals. Therefore, an OFDM transmitter can be implemented by merely substituting the Heisenberg transform with the IFFT. Similarly, the Wigner transform is replaced by the FFT operation at the receiver side. Using the mentioned transmitter and receiver structures, as well as the time-varying channel property, the OFDM DD domain channel can be modeled as $\mathbf{H}_{pq}^\mathsf{ofdm}=\sum_{i=1}^{L_{pq}}\mathbf{h}_{pq,i}^T\otimes[\mathbf{F}_M\mathbf{Q}_{pq}^{(i)}{ \mathbf{F}}_M^{\dagger}]\in\mathbb{C}^{M\times M_{\mathrm t}M}$ with $\mathbf{Q}_{pq}^{(i)}$ given in~\cite{raviteja2018interference} as
\begin{align}&\hspace {-0.5pc}\big [{\mathbf{Q}}_{pq}^{(i)}\big]_{(m,n)} = \delta \left ({\left ({m-n-\frac {\tau _{pq,i}M}{T}}\right)_{M}}\right) e^{\frac {j2\pi }{M}(n-1)\nu _{pq,i}}, \nonumber \\&\qquad\qquad\qquad\qquad\qquad\qquad\qquad m,n=1,\ldots,M,\end{align}

\noindent where $\delta(\cdot)$ represents the Dirac delta function. 

In OFDM-signal-based systems, there are two primary channel estimation approaches: one involves inserting pilot sequences across all OFDM subcarriers over a time $D_{t}$, referred to as a block-type (BT) scheme, and the other periodically inserts pilot sequences throughout the OFDM symbols with a frequency interval $D_{f}$, which is known as a comb-type scheme. It should be noted that intervals $D_{t}$ and $D_{f}$ must satisfy the conditions of $D_{t}<\frac{1}{2\nu_{\max}T}$ and $D_f<\frac1{\Delta f\tau_{\max}}$, respectively.

After obtaining estimate $\hat{\mathbf{H}}_{pq}^{\mathsf{ofdm}}$ of channel matrix $\mathbf{H}_{pq}^\mathsf{ofdm}$, the transmitting APs use MR precoding to transmit downlink ISAC signals. The $n$th transmitted OFDM symbol from the $p$th transmitting AP is expressed by
\begin{equation}
\bar{\mathbf{x}}_{p}(n)=\sqrt{\rho_{\mathrm{d}}}\sum_{q=0}^{K_{\mathrm{u}}}\bar{\eta}_{pq}^{1/2}\hat{\mathbf{H}}_{pq}^{\mathsf{ofdm}^\dagger}\bar{\mathbf{s}}_q(n),
\end{equation}

\noindent where $\bar{\mathbf{s}}_q(n)$ is the $n$th intended OFDM symbol for the $q$th user; $\bar{\eta}_{pq}$, where $p=1,\ldots,N_{\mathrm{tx}}$ and $q=0,\ldots,K_{\mathrm{u}}$, is the power control coefficient introduced to make each transmitting AP satisfy the power constraint of $\mathbb{E}\left\{\|\bar{\mathbf{x}}_{p}(n)\|^2\right\}\leq M\rho_{\mathrm{d}}$.

Similar to \eqref{eq:z_q}, the OFDM signal received by the $q$th user in the DD domain is given as
\begin{equation}
\bar{\mathbf{z}}_{q}(n)=\sqrt{\rho_{\mathrm{d}}}\sum_{p=1}^{N_{\mathrm{tx}}}\sum_{q^{\prime}=0}^{K_{\mathrm{u}}}\bar{\eta}_{pq^{\prime}}^{1/2}\mathbf{H}_{pq}^{\mathsf{ofdm}}\hat{\mathbf{H}}_{pq^{\prime}}^{\mathsf{ofdm}^\dagger}\bar{\mathbf{s}}_{q^{\prime}}(n)+\bar{\mathbf{w}}_{q}(n),
\end{equation}

\noindent where $\bar{\mathbf{w}}_{q}(n)\in\mathbb{C}^{M\times1}$ is the noise vector at the $q$th user. 

The properties of $\bar{\mathbf{Q}}_{pq}^{(i)}=\mathbf{F}_M\mathbf{Q}_{pq}^{(i)}{ \mathbf{F}}_M^{\dagger}\in\mathbb{C}^{M\times M}$ can be confirmed to be similar to those of ${\mathbf{T}} _{pq}^{(i)}$~\cite{mohammadi2022cell}. Consequently, the achievable downlink SE for the OFDM-signal-based system can be constructed by applying a similar procedure as in Theorem~\ref{thm1}. For a fair comparison, we assume that each OFDM symbol duration is $T=T_{\mathrm{cp}}+T_0$, where $T_{\mathrm{cp}}$ and $T_0$ denote the CP and data symbol durations, respectively. Therefore, the corresponding pre-log factor of the OFDM-signal-based system is $\omega_{\mathsf{ofdm}}=\left(1-\frac{N_{\mathrm{cp}}}{M}\right)$~\cite{li2023cell}.

\section{Power Allocation}
\subsection{Sensing SINR}
In this study, sensing SINR is used as a sensing performance metric of the CF-ISAC system~\cite{liu2023joint,behdad2024joint}. After defining the sensing channel matrix by $\mathbf{V}_{pr}\triangleq\mathbf{h}_{tr}^{\phantom{T}}\mathbf{h}_{pt}^{T}\in\mathbb{C}^{M_\mathrm{t}\times M_\mathrm{t}}$, the result of the sensing SINR can be obtained as follows.

\begin{lemma}\label{lem2}
The sensing SINR is given by
\begin{small}
  \begin{align}
    &\textnormal{SINR}^{(\mathrm s)}=\nonumber \\
    &\quad\!\frac{ \rho_{\mathrm{d}}\sum\limits_{r=1}^{N_{\mathrm{rx}}}{\sum\limits_{p=1}^{N_{\mathrm{tx}}}{\sum\limits_{q=0}^{K_{\mathrm{u}}}{\sum\limits_{i=1}^{L_{pq}}{\eta _{pq}\sigma_{pr}^{2}\beta_{pr}\mathrm{Tr}\left( \mathbf{V}_{pr}^{\phantom{\dagger}}\mathbf{B}_{pq,i}\mathbf{V}_{pr}^{\dagger} \right) }}}}}{ \rho_{\mathrm{d}}\sum\limits_{r=1}^{N_{\mathrm{rx}}}{\sum\limits_{p=1}^{N_{\mathrm{tx}}}{\sum\limits_{q=0}^{K_{\mathrm{u}}}{\sum\limits_{i=1}^{L_{pq}}{\eta _{pq}\mathrm{Tr}\left( \mathbf{R}_{\mathrm{rx},(pr)} \right)\mathrm{Tr}\left( \mathbf{R}_{\mathrm{tx},(pr)}\mathbf{B}_{pq,i} \right) }}}}\! +\! N_{\mathrm{rx}}M_\mathrm{t} }.
  \label{eq:sensing_SINR}
  \end{align}
\end{small}
\end{lemma}

\begin{IEEEproof}
  The proof is given in Appendix~\ref{app:lem2}.
\end{IEEEproof} 

\subsection{Power Allocation for ISAC}
This section maximizes the communication SINR, denoted by $\text{SINR}^{(\mathrm c)}_{q}$, under the assumption that the target exists. The max-min fairness optimization problem can be defined as follows:
\begin{subequations}\label{eq:35}
\begin{alignat}{2}
& \underset{\boldsymbol{\eta}\geq\mathbf{0}}{\text{maximize}} &\quad& \underset{q\in\{1,...,K_{\mathrm u}\}}{\text{min}}\Big\{\text{SINR}^{(\mathrm c)}_{q}\Big\} \label{eq:35a}
\\ 
& \text{subject to} && \text{SINR}^{(\mathrm s)} \geq \gamma_{\mathrm s} \label{eq:35b}
\\
& && \rho_{p}\leq \rho_{\mathrm{d}},\quad p=1,\ldots,N_\mathrm{tx}, \label{eq:35c}
\end{alignat}
\end{subequations}

\noindent where $\boldsymbol{\eta}\triangleq[\eta_{10}\ldots\eta_{1K_{\mathrm{u}}}\ldots\eta_{N_{\mathrm{tx}}K_{\mathrm{u}}}]^T\in\mathbb{R}^{N_{\mathrm{tx}}(K_{\mathrm{u}}+1)\times1}$, $\gamma_{\mathrm s}$ is the minimum required sensing SINR threshold, and $\rho_{p}$ is the normalized downlink SNR at the $p$th transmitting AP defined by~\eqref{eq:power_constraint}. 

\begin{remark}\label{remark2}
  During the power allocation process, an actual channel $\mathbf{R}_{pq,i}$ in the optimization problem is unknown at transmitting APs. In view of that, an estimated channel $\mathbf{B}_{pq,i}$ is used as a substitute to solve the optimization problem. In the numerical result section, the achievable downlink SE under the assumption that the transmitting APs have perfect CSI for downlink precoding and solving the optimization problem is also provided as a benchmark.
\end{remark}

Using the context of remark \ref{remark2}, the objective function $\text{SINR}^{(\mathrm c)}_{q}$ in \eqref{eq:DL_SE} can be rewritten as
\begin{small}
\begin{align}\text{SINR}^{(\mathrm c)}_{q}\! &= \frac{\rho_{\mathrm{d}}\left( \sum_{p=1}^{N_{\mathrm{tx}}}{\eta _{pq}^{1/2}}{\sum_{i=1}^{L_{pq}}}\mathrm{Tr}\left( \mathbf{B}_{pq,i} \right) \right) ^2}{\rho_{\mathrm{d}}\sum_{p=1}^{N_{\mathrm{tx}}}\sum_{q^{\prime}=0}^{K_{\mathrm{u}}}{\eta _{pq^{\prime}}}\!\!\left( \sum_{i=1}^{L_{pq}}\sum_{j=1}^{L_{pq}}{\mathrm{Tr}\left(\mathbf{B}_{pq,i}\mathbf{B}_{pq^{\prime}\!,j} \right)}\!\right)\!+\!1} \nonumber\\
&= \frac{\rho_{\mathrm{d}}\left( \sum_{p=1}^{N_{\mathrm{tx}}}{\eta _{pq}^{1/2}}b_{pq}\right) ^2}{\rho_{\mathrm{d}}\sum_{p=1}^{N_{\mathrm{tx}}}\sum_{q^{\prime}=0}^{K_{\mathrm{u}}}{\eta _{pq^{\prime}}}a_{pq,q^{\prime}}+1},\label{eq:SINR_c}\end{align}
\end{small}

\noindent where $a_{pq,q^{\prime}}\triangleq\sum_{i=1}^{L_{pq}}\sum_{j=1}^{L_{pq}}{\mathrm{Tr}\left(\mathbf{B}_{pq,i}\mathbf{B}_{pq^{\prime},j}\right)}$ and $b_{pq}\triangleq{\sum_{i=1}^{L_{pq}}}\mathrm{Tr}\left( \mathbf{B}_{pq,i}\right)$ are defined.

\begin{lemma}\label{lem3}
The power constraint in~\eqref{eq:35c} at the $p$th transmitting AP can be expressed by
\begin{equation}\label{eq:power_constraint_new}
\sum _{q=0}^{K_{\mathrm{u}}}\sum _{i=1}^{ L_{pq}} \eta _{pq} \mathrm{Tr}(\mathbf{B}_{pq,i})=\sum _{q=0}^{K_{\mathrm{u}}}\eta _{pq}b_{pq}\leq 1. 
\end{equation}
\end{lemma}

\begin{IEEEproof}
  The proof is given in Appendix~\ref{app:lem3}.
\end{IEEEproof} 

Further, it can be readily checked that the objective function in~\eqref{eq:SINR_c} has a fractional programming form, whereas the remaining two constraints \eqref{eq:35b}, \eqref{eq:power_constraint_new} can be converted into affine inequalities. Therefore, by performing the quadratic transform~\cite{shen2018fractional}, the original optimization problem~\eqref{eq:35} can be reformulated as a convex optimization problem, which can be expressed as
\vspace{-2.5pt}
\begingroup\makeatletter\def\f@size{9}\check@mathfonts
\begin{subequations}
\begin{alignat}{2}
& \underset{\boldsymbol{\eta}\geq\mathbf{0},~z}{\text{maximize}} &\quad& z 
\\ 
& \text{subject to} && \text{NUM}^{(\mathrm s)} \geq \gamma_{\mathrm s}\text{DEN}^{(\mathrm s)} 
\\
& && \sum _{q=0}^{K_{\mathrm{u}}}\eta _{pq}b_{pq}\leq 1,\quad p=1,\ldots,N_\mathrm{tx}
\\
& && 2y_q\sqrt{\rho_{\mathrm{d}}}\sum_{p=1}^{N_{\mathrm{tx}}}\eta _{pq}^{1/2}b_{pq}-y_q^2\bigg(\rho_{\mathrm{d}}\sum_{p=1}^{N_{\mathrm{tx}}}\sum_{q^{\prime}=0}^{K_{\mathrm{u}}}{\eta _{pq^{\prime}}}a_{pq,q^{\prime}}+1\bigg)\!,\nonumber\\
& && \qquad\qquad\qquad\qquad\qquad\qquad q = 1, ..., K_{\mathrm u},
\end{alignat}
\label{eq:opt_main}
\end{subequations}
\endgroup
\vspace{-12.5pt}

\noindent where $\text{NUM}^{(\mathrm s)}$ and $\text{DEN}^{(\mathrm s)}$ are the numerator and denominator of $\text{SINR}^{(\mathrm s)}$ in~\eqref{eq:sensing_SINR}, respectively, with the auxiliary variable $y_q$ for fixed $\boldsymbol{\eta}$ given by
\begin{equation} 
y_q=\frac{\sqrt{\rho_{\mathrm{d}}}\sum _{p=1}^{N_{\mathrm{tx}}}\eta_{pq}^{1/2}b_{pq}}{\rho_{\mathrm{d}}\sum _{q^{\prime}=0}^{K_{\mathrm{u}}}\sum _{p=1}^{N_{\mathrm{tx}}}\eta_{pq}a_{pq,q^{\prime}}+1}.
\label{eq:opt_yq}
\end{equation}

\noindent The above problem can be solved using the iterative approach, whose steps are outlined in Algorithm~\ref{alg1}. Under some mild conditions, this algorithm converges to the globally optimal solution of the original problem~\eqref{eq:35}~\cite{shen2018fractional}.

\begin{algorithm}[t]
\caption{Iterative Algorithm for ISAC Power Allocation}
\label{alg1}
\begin{algorithmic}[1]
\STATE $\textbf{Initialization:}$ Given an arbitrary initial positive $\boldsymbol{\eta}^{(0)}$ and the tolerance $\epsilon>0$. Set the iteration counter to $t=0$ and $z^{(0)}=0$.
\REPEAT
\STATE $t\leftarrow t+1$.
    \STATE Update $y_q^{(t)}$ according to \eqref{eq:opt_yq};
    \STATE Update $\boldsymbol{\eta}^{(t)}$ by solving the convex optimization problem \eqref{eq:opt_main} for fixed $y_q$;
\UNTIL $|z^{(t)}-z^{(t-1)}|\leq\epsilon$.
\STATE $\textbf{Output:}$ The transmit power coefficients $\boldsymbol{\eta}^{(t)}$.
\end{algorithmic}
\end{algorithm}

\section{Numerical Results}
\label{sec:Numerical Results}

In this section, numerical results are presented to provide deeper insights into the application of the OTFS signals to the CF-ISAC systems. The effectiveness of the proposed power allocation algorithm is examined, and comparisons are made with the OFDM-signal-based system. The key simulation parameters are set referring to the relevant literature~\cite{behdad2024multi,mohammadi2022cell}, and they are outlined in Table~\ref{tab3} unless otherwise specified.

\begin{table}[!t]
\centering
\caption{Simulation Parameters}
\label{tab3}
\rowcolors{2}{lightgray!30}{white}
\renewcommand\arraystretch{1.5}{
    \centering
    \begin{tabular}{ccc}
    \Xhline{0.8pt}\rowcolor{gray!35}
        \textbf{Parameters} & \textbf{Symbol} & \textbf{Value} \\ \Xhline{0.5pt}
        Carrier frequency & $f_{c}$ & 4\,GHz \\  
        Subcarrier bandwidth & $\Delta f$ & 15\,kHz \\ 
        Number of subcarriers & $M$ & 512 \\  
        Number of symbols & $N$ & 128 \\  
        Scenario size & - & 1km\,$\times$\,1km \\   
        Number of paths & $L_{pq}$ & 9 \\ 
        Number of transmitting APs & $N_{\mathrm{tx}}$ & 100 \\  
        Number of receiving APs & $N_{\mathrm{rx}}$ & 2 \\ 
        Number of antennas at each AP & $M_{\mathrm t}$ & 4 \\  
        Number of users & $K_{\mathrm u}$ & 15 \\ 
        Maximum moving speed $($UE$/$Target$)$ & $v_{\max}$ & 300\,km/h \\  
        Maximum delay spread (EVA) & $\tau_{\max}$& 2.5\,$\mu s$ \\ 
        CP sample length & $N_{\mathrm{cp}}$ & $\lceil\tau_{\max}M\Delta f\rceil$ \\ 
        Sensing SINR threshold & $\gamma_{\mathrm s}$ & 3\,dB \\  
        Noise figure & $F$ & 7\,dB \\ 
        Scaling parameter & $\mathfrak{s}$ & 0.3 \\  
        RCS variance & $\sigma_{rcs}^2$ & 0\,dBsm \\ 
        Uplink power control coefficient & $\eta_q$ & 1 (full power) \\ 
        Downlink power control coefficient & $\eta_{pq}$ & Optimum \\ 
        Channel model & - & 3GPP UMi \\ \Xhline{0.8pt}
    \end{tabular}}
\end{table}

The path loss for the communication and target-free channels are modeled by the 3GPP Urban Microcell (UMi) model defined in~\cite[Table 7.4.1-1]{3gpp2017study}, with the difference between them being that the latter channel gains are multiplied by a scaling parameter, $\mathfrak{s}\in(0,1)$, to suppress the known parts of the target-free channels attributed to LOS and permanent obstacles~\cite{behdad2024multi}. The antenna gains are set to $G_p=G_r=0\,\text{dBi}$ and the noise variance is configured by $\sigma_n^2=k_BT_0(M\Delta f)F$~W, where $k_B=1.381\times10^{-23}\,\,\text{Joules/K}$, $T_0=290$\,K, and $F=7$\,dB are the Boltzmann constant, noise temperature, and noise figure, respectively. For each tap of the channel realization between the $p$th transmitting AP and $q$th user, the delay index and Doppler index are randomly selected according to the uniform distribution; that is, $-k_{\max}\leq k_{pq,i}\leq k_{\max}$, $0\leq \ell_{pq,i}\leq \ell_{\max}$, $\forall p,q,1\leq i\leq L_{pq}$~\cite{li2021performance}.

A total region of 1\,km $\times$ 1\,km is considered within which $N_\mathrm{tx}+N_{\mathrm{rx}}$ APs and $K_\mathrm{u}$ users are randomly generated. During each sensing period, the target's location is fixed, but it can be anywhere within a 15\,m $\times$ 15\,m sensing hotspot area located in the center of the region. The $N_{\mathrm{rx}}=2$ closest APs to the sensing hotspot area are designated as sensing receivers, and the remaining $N_\mathrm{tx}$ APs serve as ISAC transmitters~\cite{behdad2024multi}. The maximum transmit power per AP is 1\,W, and the uplink pilot transmission power is 0.2\,W. For simplicity, it is assumed that the RCS values are independent and have the same variance $\sigma_{pr}^2=\sigma_{rcs}^2$ for each pair consisting of a transmitting AP $p$ and a receiving AP $r$.

This study compares four benchmark system designs: i)~the OTFS with the EP-based channel estimation but without optimization of power allocation (\emph{OTFS, Equal, EP}); ii)~the OTFS with the EP-based channel estimation and optimization Algorithm~\ref{alg1} (\emph{OTFS, Optimum, EP}); iii)~the OFDM with the BT-based channel estimation but without optimization (\emph{OFDM, Equal, BT}); iv)~the OFDM with the BT-based channel estimation and optimization (\emph{OFDM, Optimum, BT}). These system designs are also evaluated with their \emph{perfect CSI} counterparts. In the cases without optimization, the APs transmit with \emph{equal} power, and the power control coefficients at the $p$th transmitting AP are $\eta_{pq}=\left(\sum_{q'=1}^{K_{u}}\sum_{i=1}^{L_{pq}}\mathrm{Tr}\left(\mathbf{B}_{pq^{\prime},i} \right)\right)^{-1},\forall q=1,\ldots,K_{u}$. In addition, all four downlink SE components, namely the desired signal (\emph{DS}), precoding gain uncertainty (\emph{BU}), inter-symbol interference (\emph{ISI}), and inter-user interference (\emph{IUI}), in~\eqref{eq:received_DL_signal} are investigated to reveal the inhibitory influence of the optimization on the interference components.

\begin{figure}[!t]
    \centering
    \includegraphics[width=3 in]{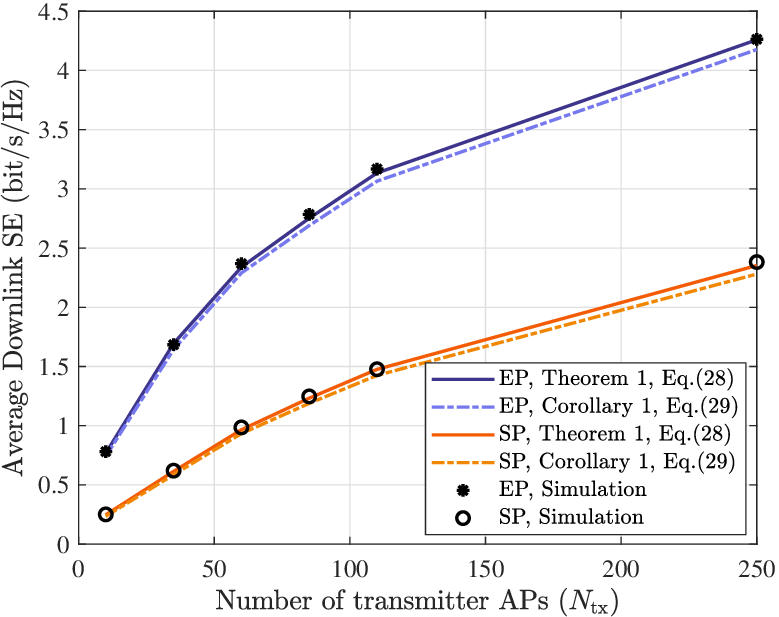}
    \caption{The average per-user downlink SE versus the number of transmitting APs.}
    \label{fig:fig_2}
\end{figure}

Fig.~\ref{fig:fig_2} presents the analytical average per-user SE in Theorem~\ref{thm1} and the lower bound in Corollary~\ref{cor1}, along with the corresponding simulated results. The OTFS frame consists of $N=8$ symbols and $M=16$ subcarriers, with the maximum delay spread of $\tau_{\max}= 5\,\mu s$. For a more comprehensive analysis, a comparison with the superimposed pilot (\emph{SP})-based channel estimation scheme~\cite{mishra2021otfs} is also conducted. The key observations are as follows. First, the proposed lower bound aligns closely with the original SE expression in~\eqref{eq:DL_SE}, while providing lower complexity for SE computation, thus paving the way for the CF-ISAC system performance optimization. Second, the \emph{SP}-based channel estimation is outperformed by the \emph{EP}-aided channel estimation. This phenomenon is due to the interference caused by data signals boosted in the absence of the guard region.

\begin{figure}[!t]
    \centering
    \includegraphics[width=3 in]{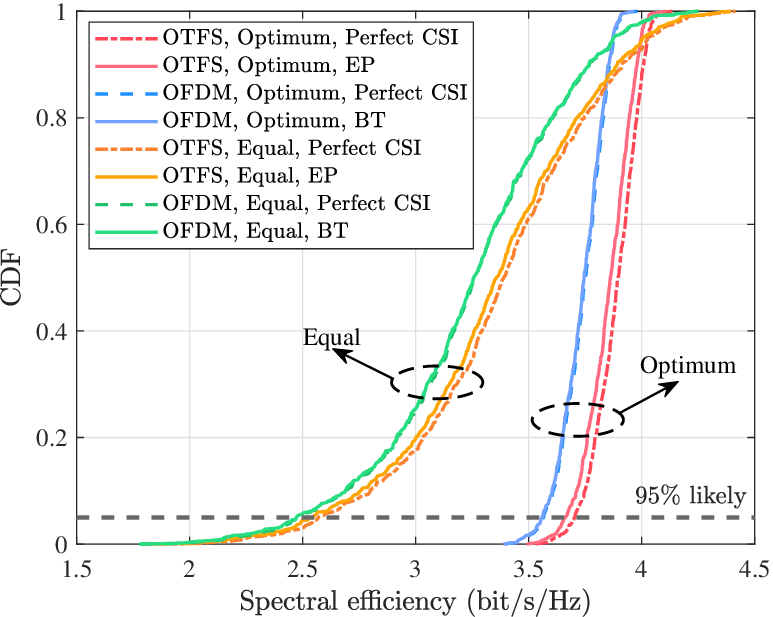}
    \caption{The cumulative distribution of the per-user downlink SE with and without optimization.}
    \label{fig:fig_3}
\end{figure}

The cumulative distributions of the per-user downlink SE both with and without optimization in the absence of sensing consideration (i.e., $\gamma_{\mathrm s}\approx 0$) are presented in Fig.~\ref{fig:fig_3}, where it can be observed that the \emph{optimum} max-min fairness-based power allocation yields a 1.45 fold higher 95\%-likely achievable SE compared to the \emph{equal} power allocation. In addition, there is a minimal disparity between its lower and upper tails, both of which denote anticipated properties of the CF-mMIMO architecture. Further, it can be noted that regardless of whether the OTFS or the OFDM signal is used, the gap between the achievable SE attained by the channel estimation and the perfect CSI is considerably smaller in comparison to the performance gap between the OTFS and the OFDM. For the OTFS-EP method, this finding can be attributed to the significant role that the guard region plays in channel estimation. However, for the OFDM-BT method, this can be caused by the fact that orthogonal pilots can be assigned to the majority of the 15 users under the 15\,kHz subcarrier bandwidth, which could lead to minimal pilot contamination, thus enhancing the channel estimation performance. The effect of the subcarrier bandwidth, along with the other potential factors on the performance difference between the OTFS and the OFDM signals, will be presented in what follows.

\begin{figure}[!t]
	\centering
	\includegraphics[width=3 in]{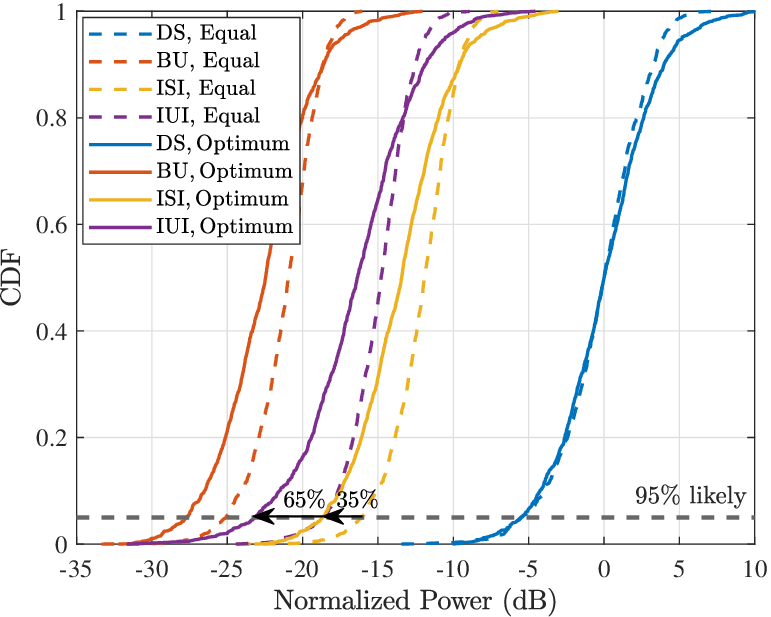}
	\caption{The cumulative distribution of normalized power components in the downlink SE with and without optimization.}
	\label{fig:fig_4}
\end{figure}

Next, diverse power components affecting the downlink SE of the OTFS-signal-based system are analyzed. All four SE components are normalized by the median of the \emph{DS}, as shown in Fig.~\ref{fig:fig_4}. The initial observation is that regardless of the status (i.e., with or without optimization), the \emph{ISI} predominates among the three interference components, exerting the most significant influence on the achievable SE. When the optimization is performed, the three interference types experience degrees of suppression based on the 95\%-likely evaluation criterion. However, the \emph{IUI} exhibits the largest decrease, reaching 65\%. This substantial decrease results from the optimization of the power control coefficients at the frame level, which significantly limits the interference from the other users. In contrast, due to the absence of in-depth optimization at the symbol level, the reduction of the \emph{ISI} is less pronounced, reaching a value of 35\%. Furthermore, the \emph{BU}, caused by channel fluctuations, has the minimum effect on the achievable SE in the simulation settings, which can be attributed to the prolonged duration of the OTFS frame in alleviating such uncertainty~\cite{demir2021foundations}.

\begin{figure}[!t]
    \centering
    \includegraphics[width=3 in]{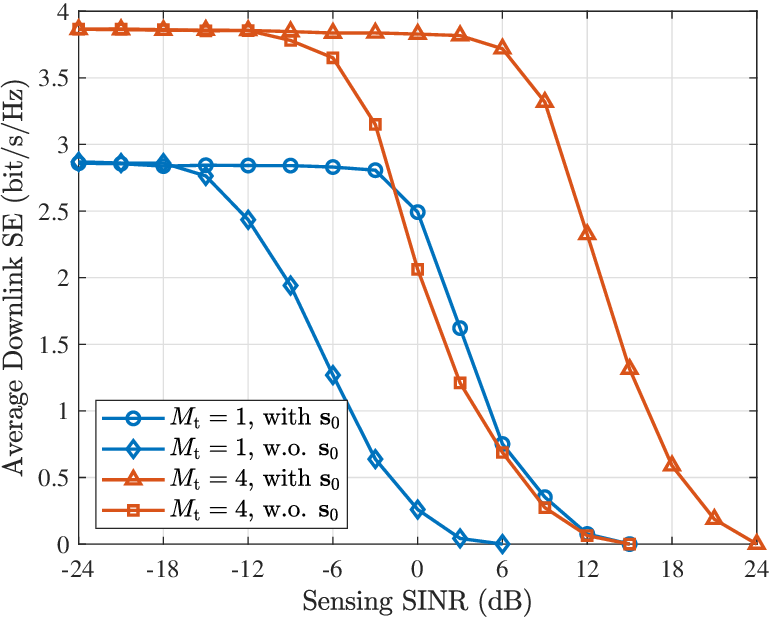}
    \caption{Tradeoff between the communication downlink SE and the sensing SINR constraint with different numbers of the AP antennas $M_{\mathrm t}=1, 4$.}
    \label{fig:fig_5}
\end{figure}

The tradeoff between the communication downlink SE and the sensing SINR constraint for different numbers of AP antennas $M_{\mathrm t}=\text{1, 4}$, both with and without (marked as w.o.) the dedicated sensing beam $\mathbf{s}_{0}$, is illustrated in Fig.~\ref{fig:fig_5}. The results indicate that an increase in the sensing SINR constraints corresponds to a decrease in average downlink SE. Particularly, for the $M_{\mathrm t}=\text{4}$ scenario, when the sensing SINR constraint is below -12\,dB, this constraint does not affect the downlink SE of the communication, which is fixed at approximately 3.8\,bit/s/Hz. Therefore, this section of the tradeoff boundary is communication-constrained, which implies that the users' communication signals are enough to satisfy the sensing constraint. Continual increments in the sensing SINR constraint shift the tradeoff boundary into the dual-constrained section, and eventually, when the sensing SINR constraint is above 18\,dB, the corresponding problem becomes infeasible. Another remarkable finding is that with the dedicated sensing beam, the downlink SE of the single-antenna scenario achieves a similar level to the $M_{\mathrm t}=\text{4}$ scenario without the dedicated sensing beam in the dual-constrained section, despite a far smaller number of used antennas. This enhancement results from the extra degrees of freedom that the dedicated beam offers during the optimization, which provides a higher optimization feasibility and strengthens the ISAC system's robustness.

\begin{figure}[!t]
    \centering
    \includegraphics[width=3 in]{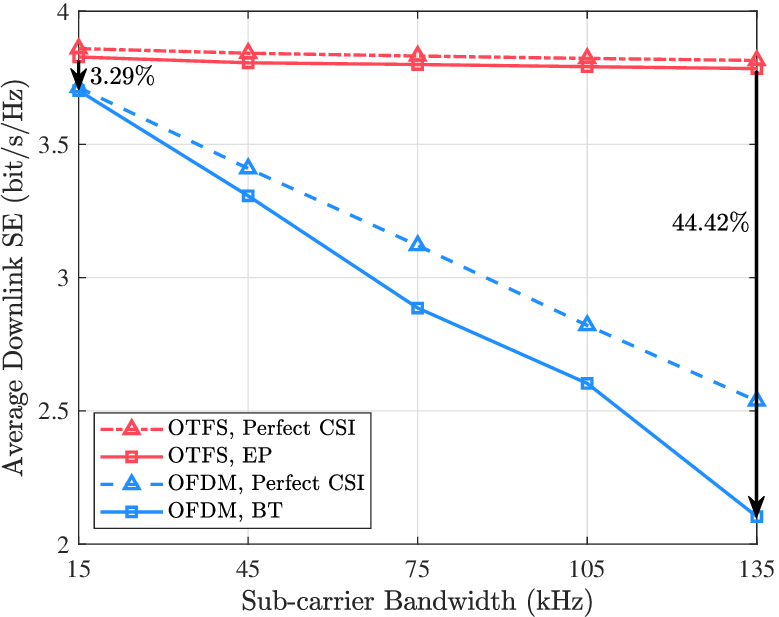}
    \caption{The average per-user downlink SE versus the subcarrier bandwidth.}
    \label{fig:fig_6}
\end{figure}

\begin{figure}[!t]
    \centering
    \includegraphics[width=3 in]{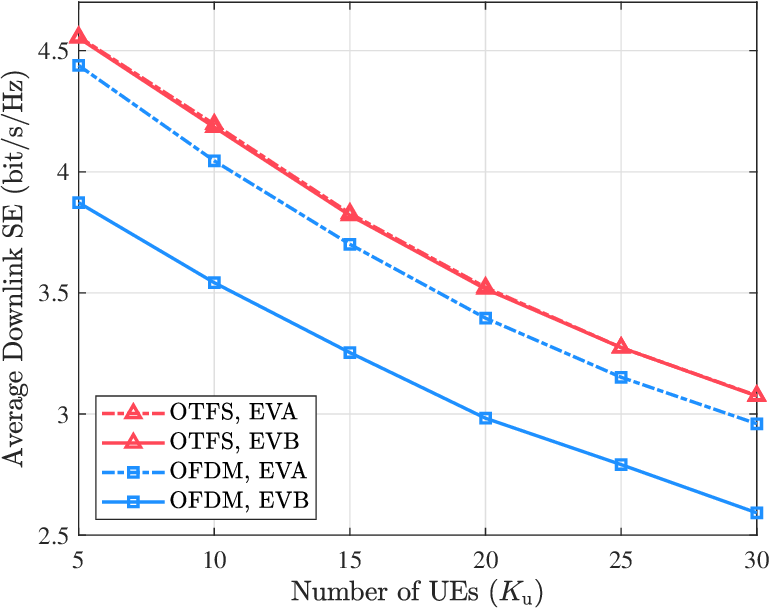}
    \caption{The average per-user downlink SE versus the number of users for the EVA and EVB model.}
    \label{fig:fig_7}
\end{figure}

The performance gap between the systems based on the OTFS and OFDM signal across various subcarrier bandwidths is shown in Fig.~\ref{fig:fig_6}. The performance gap increases up to 13 times when the subcarrier bandwidth increases from 15\,kHz to 135\,kHz. The impact of this on the achievable SE of the OFDM-signal-based system is two-fold. First, the growth in the subcarrier bandwidth leads to a substantial increase in the CP overhead of the OFDM signal compared to the OTFS, which requires merely a single guard interval between two OTFS blocks to prevent inter-block interference. Second, fewer subcarriers can be accommodated within a specific coherent bandwidth, thus shortening the OFDM pilot sequence and intensifying the pilot contamination — a deficiency not inherent to the OTFS. Although the increase in the system bandwidth elevates the received thermal noise, its adverse effect results in only a slight decrease in the SE performance.

\begin{table}[!t]
\centering
\caption{Key gap factors in OTFS and OFDM-signal-based systems}
\label{tab4}
\begin{threeparttable}
\renewcommand\arraystretch{1.5}{
    \centering
    \begin{tabular}{|c|c|c|c|}
    \hline
        \makecell{\textbf{Subcarrier} \\ \textbf{bandwidth}} & \makecell{\textbf{CP overhead of} \\ \textbf{OTFS signal} \\ \textbf{EVA (EVB)}} & \makecell{\textbf{CP overhead of} \\ \textbf{OFDM signal} \\ \textbf{EVA (EVB)}} & \makecell{\textbf{Pilot length of} \\ \textbf{OFDM signal}} \\ \hline
        15\,kHz & 0.03\% (0.12\%)  & ~3.91\%~~(15.04\%) & 14 \\ \hline
        45\,kHz & 0.09\% (0.35\%)  & 11.33\% (45.12\%) & 4 \\ \hline
        75\,kHz & 0.15\% (0.59\%)  & 18.75\% (75.00\%) & 2 \\ \hline
        105\,kHz & 0.21\% (0.82\%) & \multicolumn{1}{l|}{\,26.37\% \quad (\ding{56})*}  & 2 \\ \hline
        135\,kHz & 0.26\% (1.06\%) & \multicolumn{1}{l|}{\,33.79\% \quad (\ding{56}) }  & 1 \\ \hline
    \end{tabular}}
    \begin{tablenotes}
    \item[*] The maximum delay is so large that it will inevitably be distributed by the previous frame.
	\end{tablenotes}
\end{threeparttable}
\end{table}

Finally, a comparative study of the extended vehicular A (EVA) model and the extended vehicular B (EVB) model is conducted, with a reference to the number of users, as represented in Fig.~\ref{fig:fig_7}. The distinguishing parameters of the EVB model are $L_{pq}=6$ and $\tau_{\max}=10\,\,\mu\text{s}$. This study exposes an additional advantage of using the OTFS over the OFDM signal; namely, the OTFS can maintain stable ISAC performance in spite of the increase in the channel's maximum delay $\tau_{\max}$. Unlike the significant increase in the CP overhead in the OFDM-signal-based system, the CP overhead of the OTFS-signal-based system shows only a marginal increase. Specific impact factor values for Fig.~\ref{fig:fig_6} and Fig.~\ref{fig:fig_7} are presented in Table~\ref{tab4}.

\section{Conclusion}

This paper studies the implementation of the emerging OTFS signal in the CF-ISAC systems, concurrently optimizing and assessing the system's overall performance. A closed-form SE expression of the downlink CF-ISAC system with multi-antenna AP is derived, and a low-complexity achievable SE lower bound is defined, which is applicable to all OTFS-signal-based systems. Next, an iterative power allocation algorithm is proposed to maximize the lowest SE of users while satisfying the constraints of both sensing SINR and per-AP power to improve the system's performance. The numerical results indicate the tightness of the proposed lower bound, which paves the way for performing the power allocation algorithm and analyzing the impact of the SE expression's components. In addition, the vital role of the dedicated sensing beam in elevating the ISAC performance is elucidated, providing an innovative strategy to improve the robustness of ISAC systems. Moreover, the superiority of the OTFS signal over the OFDM signal in terms of the system's achievable SE is demonstrated. Particularly, a remarkable 13-fold gain in the SE performance gap is observed with the subcarrier bandwidth expansion. Contrary to the OFDM signal, using the OTFS signal demonstrates a stable SE across different maximum delay spreads. These promising properties of the OTFS signal could encourage its future application, particularly in the field of millimeter waves with a large bandwidth.

{\appendices
\section{Proof of Lemma \ref{lem1}}
\label{app:lem1}

Based on the received pilot signal in~\eqref{eq:received_pilot}, the MMSE channel estimation can be performed. According to the estimation theory~\cite[Cor. 4.1]{demir2021foundations,li2023cell}, the MMSE estimator can be defined as follows:
\begingroup\makeatletter\def\f@size{9}\check@mathfonts
  \begin{equation}
    \hat{\mathbf{h}}_{pq,i}=\frac{\mathbb{E}\left\{\mathbf{h}_{pq,i} \big( \mathbf{y}_p\left[ k,\ell \right] \big)^{\dagger}\right\}}{\mathbb{E}\left\{\mathbf{y}_p\left[ k,\ell \right] \big( \mathbf{y}_p\left[ k,\ell \right] \big)^{\dagger}\right\}} \mathbf{y}_p\left[ k,\ell \right]. \label{eq:MMSE}
  \end{equation}
\endgroup

Using~\eqref{eq:received_pilot}, the expectation $\mathbb{E}\{\mathbf{h}_{pq,i}\left( \mathbf{y}_p\left[ k,\ell \right] \right)^{\dagger}\}$ in the numerator can be calculated by
\begingroup\makeatletter\def\f@size{9}\check@mathfonts
    \begin{align}
      \mathbb{E}&\left\{\mathbf{h}_{pq,i}\big( \mathbf{y}_p\left[ k,\ell \right] \big)^{\dagger}\right\}\nonumber\\
      &=\sqrt{\eta _{q}P_{q}^{\mathrm{Pil}}} \mathbb{E}\left\{ \mathbf{h}_{pq,i}\left( \mathbf{h}_{pq,i} \right)^{\dagger} \right\}+\underbrace{\mathbb{E}\left\{ \mathbf{h}_{pq,i}\big[ \boldsymbol{\mathcal{I}}_{1}(k,\ell) \big]^{\dagger}\right\}}_{=0}\nonumber\\
      &\qquad\quad+\underbrace{\mathbb{E}\left\{ \mathbf{h}_{pq,i}\big[ \boldsymbol{\mathcal{I}}_{2}(k,\ell) \big]^{\dagger}\right\}+\mathbb{E}\left\{ \mathbf{h}_{pq,i}\big[ \mathbf{w}_{p}[k,\ell] \big]^{\dagger}\right\}}_{=0}\nonumber\\
      &=\sqrt{\eta _{q}P_{q}^{\mathrm{Pil}}}\mathbf{R}_{pq,i},
    \label{eq:estimated_num}
    \end{align}
\endgroup

\noindent where the first equality holds for the path indicator $\tilde {b}[\ell -\ell _{q}]\neq0$, and the second equality holds because data symbols and noise are zero-mean random variables (RVs). 

Further, using~\eqref{eq:received_pilot}, the expectation $\mathbb{E}\{\mathbf{y}_p\left[ k,\ell \right] \left( \mathbf{y}_p\left[ k,\ell \right] \right)^{\dagger}\}$ in the denominator of the MMSE estimator in (\ref{eq:MMSE}) can be expressed by
\begingroup\makeatletter\def\f@size{9}\check@mathfonts
\begin{align}
    \mathbf{\Psi}_{pq,i} &\triangleq \mathbb{E}\left\{\mathbf{y}_p\left[ k,\ell \right] \big( \mathbf{y}_p\left[ k,\ell \right] \big)^{\dagger}\right\} \nonumber\\
    &=\eta _{q}P_{q}^{\mathrm{Pil}}\mathbb{E}\left\{ \mathbf{h}_{pq,i}\left( \mathbf{h}_{pq,i} \right)^{\dagger} \right\}+ \mathbb {E} \{|\boldsymbol{\mathcal{I}}_{1}(k,\ell)|^{2}\} \nonumber\\
    &\qquad\qquad\qquad\quad+ \mathbb{E}\{|\boldsymbol{\mathcal{I}}_{2}(k,\ell)|^{2}\}+\sigma _{n}^{2}\mathbf{I}_{M_\mathrm{t}}.
\label{eq:estimated_psi_ori}
\end{align}
\endgroup

\noindent Since the first expectation in~\eqref{eq:estimated_psi_ori} can be expressed as $\mathbb{E}\big\{ \mathbf{h}_{pq,i}\left( \mathbf{h}_{pq,i} \right)^{\dagger} \big\}=\mathbf{R}_{pq,i}$, the focus now shifts to the derivation of the remaining two expectation terms. Using~\eqref{eq:estimation_I1} and noting the independence between data symbols, as well as $\mathbb{E}\{|x_{dq}[k^{\prime},\ell^{\prime}]|^2\}=P_{q}^{\mathrm{dt}}$, $\mathbb{E} \{ \left| \boldsymbol{\mathcal{I}} _1(k,\ell ) \right|^2 \}$ can be obtained by \eqref{eq:estimation_I1_temp}, which is shown at the bottom of the page.

\begin{figure*}[hb] 
\xhrulefill {black}{1pt}
  \centering
  \begin{small}
    \begin{align}
	\mathbb{E} \left\{ \left| \boldsymbol{\mathcal{I}} _1(k,\ell ) \right|^2 \right\} &=\eta _q\sum_{k^{\prime}=0}^{k_{pq}}{\sum_{\ell ^{\prime}=0}^{\ell _{pq}}{b}}\left[ k^{\prime},\ell ^{\prime} \right] \sum_{c\notin \mathcal{K}}{\mathbb{E}}\left\{ \Big| x_{dq}\left[ \left( k-k^{\prime}+c \right) _N,\left( \ell -\ell ^{\prime} \right) _M \right] \Big|^2 \right\} \mathbb{E} \left\{ \Big| \mathbf{h}_{pq}\left[ \left( k-k^{\prime} \right) _N,\left( \ell -\ell ^{\prime} \right) _M \right] \alpha [k,\ell ,c] \Big|^2 \right\}\nonumber\\
	&=P_{q}^{\mathrm{dt}}\eta _q\sum_{k^{\prime}=0}^{k_{pq}}{\sum_{\ell ^{\prime}=0}^{\ell _{pq}}{b}}\left[ k^{\prime},\ell ^{\prime} \right] \sum_{c\notin \mathcal{K}}{\mathbb{E}}\left\{ \Big| \mathbf{h}_{pq}\left[ \left( k-k^{\prime} \right) _N,\left( \ell -\ell ^{\prime} \right) _M \right] \Big|^2 \right\} \Big|\alpha [k,\ell ,c]\Big|^2.
	\label{eq:estimation_I1_temp}
	\end{align}
   \end{small}
\end{figure*}

Due to the effect of fractional Doppler, the received pilot symbol on index $k_q$ is interfered with data symbols outside the guard region $\mathcal {K}$. It can be shown that the magnitude $|\alpha[k,\ell,c]|$ is almost constant for $\begin{aligned}k\in[k_q-k_{\max}-\hat{k},k_q+k_{\max}+\hat{k}]\end{aligned}$ and $c\notin\mathcal{K}$. In particular, when employing a rectangular window, $\big |\alpha [k,\ell,c]\big |^{2} \approx 1/{N^2}$~\cite{wei2021transmitter}, which results in 
\begin{equation} 
\sum _{c \notin \mathcal {K} }\big |\alpha [k,\ell,c]\big |^{2} \approx \frac {(N-4k_{\max}-4\hat {k}-1)}{N^{2}}.
\end{equation}

\noindent Therefore, using the property $\sum_{k^{\prime}=0}^{k_{pq}}\sum_{\ell^{\prime}=0}^{\ell_{pq}}b[k^{\prime},\ell^{\prime}]=L_{pq}$ yields
\begingroup\makeatletter\def\f@size{9}\check@mathfonts
\begin{align}
\mathbb {E}\Big \{\big |\boldsymbol{\mathcal {I}}_{1}(k,\ell)\big |^{2}\Big \}=&P_{q}^{\mathrm {dt}}\eta _{q}\sum _{c \notin \mathcal {K} }\big |\alpha [k,\ell,c]\big |^{2} \sum _{i=1}^{L_{pq}}\mathbb {E}\Big \{ \mathbf{h}_{pq,i}\left( \mathbf{h}_{pq,i} \right)^{\dagger}\Big\}\nonumber\\
\approx&P_{q}^{\mathrm {dt}}\eta _{q}\frac {(N-4k_{\max}-4\hat {k}-1)}{N^{2}} \sum _{i=1}^{ L_{pq}}\mathbf{R}_{pq,i}.
\label{eq:estimation_I1_new}
\end{align}
\endgroup

After similar steps, we obtain
\begingroup\makeatletter\def\f@size{9}\check@mathfonts
\begin{equation} 
\mathbb {E}\Big \{\big |\boldsymbol{\mathcal {I}}_{2}(k,\ell)\big |^{2}\Big \} \approx \frac {1}{N} \sum _{q'\neq q}^{K_{\mathrm{u}}} \eta _{q^{\prime} } {{P_{q'}^{\mathrm {dt}}}} \sum _{i=1}^{ L_{pq}}\mathbf{R}_{pq,i}.
\label{eq:estimation_I2_new}
\end{equation}
\endgroup

Then, by substituting~\eqref{eq:estimation_I1_new}~and~\eqref{eq:estimation_I2_new} into~\eqref{eq:estimated_psi_ori}, $\mathbf{\Psi}_{pq,i}$ can be obtained as~\eqref{eq:estimated_psi}. Finally, the estimated channel and its variance in~\eqref{eq:estimation_final} are derived by inserting~\eqref{eq:estimated_num}~and~\eqref{eq:estimated_psi} into~\eqref{eq:MMSE}.

\section{Proof of Theorem \ref{thm1}}
\label{app:thm1}

First, several lemmas from~\cite{mohammadi2022cell} about the DD domain channel in \eqref{eq:DD_channel}, which support the subsequent proofs, are introduced as follows.

\begin{lemma}\label{lem4}
For the matrix $\mathbf{T}_{pq}^{(i)}$, we have $\mathbf{T}_{pq}^{(i)} {\mathbf{T}}_{pq}^{(i)^{\dagger}} = \mathbf{I}_{MN}$. 
\end{lemma}

\begin{lemma}\label{lem5}
For two different paths with different delay indices in \eqref{eq:DD_channel}, we have
\begin{equation} \Big [{\mathbf{T}}_{pq}^{(i)} {\mathbf{T}}_{pq'}^{(j)^{\dagger} }\Big]_{(v,v)} = 0, \quad (\ell _{pq,i}-\ell _{pq,j})_{M}\neq 0.\end{equation}
\end{lemma}

\begin{lemma}\label{lem6}
For any two matrices of $\mathbf{T}_{pq}^{(i)}$ and $\mathbf{T}_{pq'}^{(j)}$ in \eqref{eq:DD_channel}, we have
\begin{equation} \Bigg |\sum _{v'=1}^{MN} \bigg [{\mathbf{T}}_{pq}^{(i)} {\mathbf{T}}_{pq'}^{(j)^{\dagger} }\bigg]_{(v,v')}\Bigg |^{2}=1.\end{equation}
\end{lemma}

As there are $MN$ sub-channels in the channel model~\eqref{eq:DD_channel}, an achievable SE at user $q$ can be defined as $R_{q}=\frac{\omega_{\mathsf{otfs}}}{MN}\sum\nolimits_{v=1}^{MN}{I_{qv}}(\text{SINR}^{(\mathrm c)}_{qv})$, where $I_{qv}(\text{SINR}_{qv})=\log_2(1+\text{SINR}^{(\mathrm c)}_{qv})$, with $\text{SINR}^{(\mathrm c)}_{qv}$; that is, the SINR at the $q$th user is expressed by~\cite{mohammadi2022cell}
\begin{align}
\text{SINR}_{qv}^{(\mathrm c)}=\frac{\left|\mathbb{DS}_{qv}\right|^2}{\mathbb{E}\{\left|\mathbb{BU}_{qv}\right|^2\}+\mathbb{E}\{\left|\mathbb{I}_{qv1}\right|^2\}+\mathbb{E}\{\left|\mathbb{I}_{qv2}\right|^2\}+1/\rho_{\mathrm{d}}}.
\label{eq:communication_SINR_ori}
\end{align}

Next, we proceed with the derivation of $\text{SINR}^{(\mathrm c)}_{qv}$. By noting that ${\mathbf{H}}_{pq} = \sum _{i=1}^{ L_{pq}} \left( \mathbf h_{pq,i}^T\otimes {\mathbf{T}} _{pq}^{(i)} \right)$ and $\hat{\mathbf{H}}_{pq} = \sum _{i=1}^{ L_{pq}} \left( \hat{\mathbf{h}}_{pq,i}^T\otimes {\mathbf{T}} _{pq}^{(i)} \right)$, the $\mathbb{DS}_{qv}$ can be derived as
\begingroup\makeatletter\def\f@size{9}\check@mathfonts
\begin{align} \mathbb {DS}_{qv}=&\sum _{p=1}^{N_{\mathrm{tx}}} {\eta }_{pq}^{1/2} \mathbb{E} \Big \{ [{\mathbf{H}}_{pq}]_{(v,:)}[\hat {\mathbf{H}}_{pq}^{\dagger}]_{(:,v)} \Big \} \nonumber\\
\stackrel {(a)}{=}&\sum _{p=1}^{N_{\mathrm{tx}}} {\eta }_{pq}^{1/2} \mathbb {E} \Bigg \{ \bigg({\sum _{i=1}^{ L_{pq}} \hat{\mathbf h}_{pq,i}^T\otimes\big [{\mathbf{T}}_{pq}^{(i)}\big]_{(v,:)} }\bigg) \nonumber
\\&\times \, \bigg({\sum _{j=1}^{ L_{pq}} \hat{\mathbf{h}}_{pq,j}^*\otimes\big [{\mathbf{T}}_{pq}^{(j)^{\dagger} }\big]_{(:,v)} }\bigg) \Bigg \}\nonumber
\\\stackrel {(b)}{=}&\sum_{p=1}^{N_{\mathrm{tx}}}{\eta _{pq}^{1/2}}\Bigg( \sum_{i=1}^{L_{pq}}{\mathbb{E}}\bigg\{ \Big( \hat{\mathbf{h}}_{pq,i}^T\hat{\mathbf{h}}_{pq,i}^* \Big) \otimes [\mathbf{T}_{pq}^{(i)}\mathbf{T}_{pq}^{(i)^{\dagger}}]_{(v,v)} \bigg\} \Bigg)  \nonumber
\\\stackrel {(c)}{=}&\sum _{p=1}^{N_{\mathrm{tx}}} {\eta }_{pq}^{1/2} {\sum _{i=1}^{ L_{pq}} \mathrm{Tr}\bigg(\mathbb {E}\left \{\hat{\mathbf{h}}_{pq,i}^*\hat{\mathbf{h}}_{pq,i}^T\right \} }\bigg) \nonumber
\\=& \sum _{p=1}^{N_{\mathrm{tx}}} \sum _{i=1}^{ L_{pq}} {\eta }_{pq}^{1/2} \mathrm{Tr}\left({\mathbf{B}_{pq,i}}\right),\label{eq:DS}\end{align}
\endgroup

\noindent where in (a), $\boldsymbol{\varepsilon}_{pq,i}=\mathbf{h}_{pq,i}-\hat{\mathbf{h}}_{pq,i}$ is substituted, and the facts $\hat{\mathbf{h}}_{pq,i}$ and $\boldsymbol{\varepsilon}_{pq,i}$ are independent RVs and $\mathbb{E}\{\boldsymbol{\varepsilon}_{pq,i}\}=\mathbf{0}$ are used; meanwhile, (b) is based on the identity $(\mathbf{A}\otimes\mathbf{B})(\mathbf{C}\otimes\mathbf{D})=(\mathbf{AC}\otimes\mathbf{BD})$ and the fact that $\hat{\mathbf{h}}_{pq,i}$ are zero-mean and independent; in (c), Lemma~\ref{lem4} is used.

Adopting the principle that the variance of a sum of independent RVs equates to the sum of individual variances, $\mathbb{E}\{\left|\mathbb{BU}_{qv}\right|^2\}$ in \eqref{eq:communication_SINR_ori} is obtained as follows:
\begingroup\makeatletter\def\f@size{9}\check@mathfonts
\begin{align}&\hspace {-0.5pc}\mathbb {E} \Big \{ \big |\mathbb {BU}_{qv}\big |^{2} \Big \} = \sum _{p=1}^{N_{\mathrm{tx}}} \eta _{pq} \Bigg (\mathbb {E} \Big \{ \Big | [{\mathbf{H}}_{pq}]_{(v,:)}[\hat {\mathbf{H}}_{pq}^{\dagger}]_{(:,v)} \Big |^{2} \Big \}\nonumber \\&\qquad\qquad\qquad\qquad- \, \Big | \mathbb {E} \Big \{ [{\mathbf{H}}_{pq}]_{(v,:)}[\hat {\mathbf{H}}_{pq}^{\dagger}]_{(:,v)} \Big \} \Big |^{2} \Bigg).\label{eq:BU_t1}\end{align}
\endgroup

Then, after recalling \eqref{eq:DD_channel} and applying Lemma~\ref{lem4}, i.e., $\Big [{\mathbf{T}}_{pq}^{(i)} {\mathbf{T}}_{pq}^{(i)^{\dagger} }\Big]_{(v,v)} = 1$, \eqref{eq:BU_t1} becomes
\begingroup\makeatletter\def\f@size{9}\check@mathfonts
\begin{equation} \mathbb {E} \Big \{ \big |\mathbb {BU}_{q}\big |^{2} \Big \} = \sum _{p=1}^{N_{\mathrm{tx}}} {\eta }_{pq} \Bigg({\mathbb {V}_{1}- \bigg(\,{\sum _{i=1}^{ L_{pq}} \mathrm{Tr}\left({\mathbf{B}_{pq,i}}\right) }\bigg)^{2} }\Bigg),\label{eq:BU_t2}\end{equation}
\endgroup

\noindent where 
\begingroup\makeatletter\def\f@size{9}\check@mathfonts
\begin{align}&\mathbb {V}_{1}= \mathbb {E} \Bigg \{ \bigg | \sum _{i=1}^{ L_{pq}} {\mathbf{h}}_{pq,i}^T\hat{\mathbf{h}}_{pq,i}^* + \sum _{i=1}^{ L_{pq}} \sum _{j\neq i}^{ L_{pq}} \big [{\mathbf{T}}_{pq}^{(i)} {\mathbf{T}}_{pq}^{(j)^{\dagger} } \big]_{(v,v)} \nonumber \\&\qquad\qquad\qquad\qquad\qquad\qquad\qquad\quad\times \, {\mathbf{h}}_{pq,i}^T\hat{\mathbf{h}}_{pq,i}^* \bigg |^{2} \Bigg \}.\end{align}
\endgroup

\begin{figure*}[hb] 
\xhrulefill {black}{1pt}
  \centering
  \begin{small}
  \begin{align}
    \mathbb{V} _1=&\mathbb{E} \left\{ \left| \sum_{i=1}^{L_{pq}}{\left( \boldsymbol{\varepsilon }_{pq,i}^T\hat{\mathbf{h}}_{pq,i}^*+\hat{\mathbf{h}}_{pq,i}^T\hat{\mathbf{h}}_{pq,i}^* \right)}+\sum_{i=1}^{L_{pq}}{\sum_{j\ne i}^{L_{pq}}{\left[ \mathbf{T}_{pq}^{(i)}\mathbf{T}_{pq}^{(j)^{\dagger}} \right] _{(v,v)}}}(\boldsymbol{\varepsilon }_{pq,i}^T+\hat{\mathbf{h}}_{pq,i}^T)\hat{\mathbf{h}}_{pq,j}^* \right|^2 \right\}\nonumber\\
    =&\mathbb{E} \left\{ \left| \sum_{i=1}^{L_{pq}}{\left( \boldsymbol{\varepsilon }_{pq,i}^T\hat{\mathbf{h}}_{pq,i}^*+\hat{\mathbf{h}}_{pq,i}^T\hat{\mathbf{h}}_{pq,i}^* \right)} \right|^2 \right\} +\mathbb{E} \left\{ \left| \sum_{i=1}^{L_{pq}}{\sum_{j\ne i}^{L_{pq}}{\left[ \mathbf{T}_{pq}^{(i)}\mathbf{T}_{pq}^{(j)^{\dagger}} \right] _{(v,v)}}}(\boldsymbol{\varepsilon }_{pq,i}^T+\hat{\mathbf{h}}_{pq,i}^T)\hat{\mathbf{h}}_{pq,j}^* \right|^2 \right\}\nonumber\\
    =&\sum_{i=1}^{L_{pq}}{\left( \mathbb{E} \left\{ \left| \boldsymbol{\varepsilon }_{pq,i}^T\hat{\mathbf{h}}_{pq,i}^* \right|^2 \right\} +\mathbb{E} \left\{ \left| \hat{\mathbf{h}}_{pq,i}^T\hat{\mathbf{h}}_{pq,i}^* \right|^2 \right\} \right)}+\sum_{i=1}^{L_{pq}}{\sum_{j\ne i}^{L_{pq}}{\mathbb{E}}}\left\{ \hat{\mathbf{h}}_{pq,i}^T\hat{\mathbf{h}}_{pq,i}^* \right\} \mathbb{E} \left\{ \hat{\mathbf{h}}_{pq,j}^T\hat{\mathbf{h}}_{pq,j}^* \right\}\nonumber\\
    &+\sum_{i=1}^{L_{pq}}{\sum_{j\ne i}^{L_{pq}}{\mathbb{E}}}\left\{ \mathfrak{R} [\Psi ] \right\} +\sum_{i=1}^{L_{pq}}{\sum_{j\ne i}^{L_{pq}}{\chi _{pq}^{ij}}}\mathrm{Tr}\left(\mathbb{E} \left\{ \boldsymbol{\varepsilon }_{pq,i}^*\boldsymbol{\varepsilon }_{pq,i}^T+\hat{\mathbf{h}}_{pq,i}^*\hat{\mathbf{h}}_{pq,i}^T \right\} \mathbb{E} \left\{ \hat{\mathbf{h}}_{pq,j}^*\hat{\mathbf{h}}_{pq,j}^T \right\}\right),
  \label{eq:BU_v1}
  \end{align}
  \end{small}
\end{figure*}

\noindent Note that $a\triangleq\sum\nolimits _{i=1}^{ L_{pq}} \sum\nolimits _{j\neq i}^{ L_{pq}}{\hat{\mathbf{h}}_{pq,i}^T}\hat{\mathbf{h}}_{pq,i}^*$ in the second term is a zero-mean RV independent of $b\triangleq\sum\nolimits_{i=1}^{L_{pq}}{\hat{\mathbf{h}}_{pq,i}^T}\hat{\mathbf{h}}_{pq,i}^*$ in the first term. Therefore, we have $\mathbb{E}\left\{|a+b|^2\right\}=\mathbb{E}\left\{|a|^2\right\}+\mathbb{E}\left\{|b|^2\right\}$. Leveraging this property, $\mathbb{V}_{1}$ can be expressed by~\eqref{eq:BU_v1}, which is shown at the bottom of the page, where $\Psi=\bigg[\Big(\big[\mathbf{T}_{pq}^{(i)}\mathbf{T}_{pq}^{(j)^\dagger}\big]_{(v,v)}\big[\mathbf{T}_{pq}^{(j)}\mathbf{T}_{pq}^{(i)^\dagger}\big]_{(v,v)}^*\Big)\big({\mathbf{h}}_{pq,i}^T\hat{\mathbf{h}}_{pq,j}^*\big)^2\bigg]$. Then, based on the facts that $\mathbb{E}\big\{\boldsymbol{\varepsilon }_{pq,i}^*\boldsymbol{\varepsilon }_{pq,i}^T\big\}=\mathbf{R}_{pq,i}-\mathbf{B}_{pq,i}$, $\mathbb{E} \left\{ \left| \hat{\mathbf{h}}_{pq,i}^T\hat{\mathbf{h}}_{pq,i}^* \right|^2 \right\} =\mathrm{Tr}^2(\mathbf{B}_{pq,i})+\mathrm{Tr}(\mathbf{B}_{pq,i}\mathbf{B}_{pq,i})$~\cite[Appendix C.3.5]{bjornson2017massive}, and $\mathbb{E}\left\{\mathfrak{Re}(\Psi)\right\}=0$, \eqref{eq:BU_v1} can be reduced to
\begingroup\makeatletter\def\f@size{9}\check@mathfonts
\begin{align}\mathbb {V}_{1}=& \sum _{i=1}^{ L_{pq}} \bigg( \mathrm{Tr}\left({\mathbf{B}_{pq,i}}\mathbf{R}_{pq,i}\right)+\mathrm{Tr}^2(\mathbf{B}_{pq,i}) \bigg) \nonumber\\
&\!\!+ \sum _{i=1}^{ L_{pq}} \sum _{j\neq i}^{ L_{pq}} \bigg({\mathrm{Tr}\left({\mathbf{B}_{pq,i}}\right)\mathrm{Tr}\left(\mathbf{B}_{pq,j}\right)+{\chi }_{pq}^{ij}\mathrm{Tr}(\mathbf{R}_{pq,i}\mathbf{B}_{pq,i}) }\bigg)\!.\label{eq:BU_v11}\end{align}
\endgroup

\noindent Next, substituting~\eqref{eq:BU_v11} into~\eqref{eq:BU_t2} yields
\begingroup\makeatletter\def\f@size{9}\check@mathfonts
\begin{align} \mathbb {E} \Big \{ \big |\mathbb {BU}_{qv}\big |^{2} \Big \} =& \sum _{p=1}^{N_{\mathrm{tx}}}\eta _{pq}\sum _{i=1}^{ L_{pq}} \Bigg(\mathrm{Tr}(\mathbf{R}_{pq,i}\mathbf{B}_{pq,i}) \nonumber\\
&\qquad\qquad+ \sum _{j\neq i}^{ L_{pq}} {\chi }_{pq}^{ij} \mathrm{Tr}(\mathbf{R}_{pq,i}\mathbf{B}_{pq,j}) \Bigg). \label{eq:BU}\end{align}
\endgroup\makeatletter\def\f@size{9}\check@mathfonts

Then, the inter-symbol interference term is derived by
\begingroup\makeatletter\def\f@size{9}\check@mathfonts
\begin{align}&\hspace {-2pc}\mathbb {E} \big \{|\mathbb {I}_{qv1}|^{2}\big \}\nonumber \\
=&\mathbb {E} \left\{{ \bigg | \sum _{p=1}^{N_{\mathrm{tx}}} \sum _{v'\neq v}^{MN} {\eta }_{pq}^{1/2} \big [{\mathbf{H}}_{pq}\big]_{(v,:)} \big [\hat {\mathbf{H}}_{pq}^{\dagger} \big]_{(:,v')} \bigg |^{2} }\right\} \nonumber \\
=&\mathbb {E} \Bigg \{ \Bigg | \sum _{p=1}^{N_{\mathrm{tx}}} \sum _{v'\neq v}^{MN} {\eta }_{pq}^{1/2} \bigg({\sum _{i=1}^{ L_{pq}} \mathbf{h}_{pq,i}^T\otimes\big [{\mathbf{T}}_{pq}^{(i)}\big]_{(v,:)}}\bigg)\nonumber \\
&\times \, \bigg({\sum _{j=1}^{ L_{pq}} \hat{\mathbf{h}}_{pq,j}^*\otimes\big [{\mathbf{T}}_{pq}^{(j)^{\dagger} }\big]_{(:,v')}}\bigg) \Bigg |^{2} \Bigg \}\nonumber \\
=&\mathbb {E} \Bigg \{ \Bigg | \sum _{p=1}^{N_{\mathrm{tx}}} \sum _{v'\neq v}^{MN} {\eta }_{pq}^{1/2} \bigg (\sum _{i=1}^{ L_{pq}} \mathbf{h}_{pq,i}^T \hat{\mathbf{h}}_{pq,i}^* \big [{\mathbf{T}}_{pq}^{(i)} {\mathbf{T}}_{pq}^{(i)^{\dagger} } \big]_{(v,v')}\nonumber \\
&+ \sum _{i=1}^{ L_{pq}} \sum _{ j\neq i}^{ L_{pq}} \mathbf{h}_{pq,i}^T \hat{\mathbf{h}}_{pq,j}^* \big [{\mathbf{T}}_{pq}^{(i)} {\mathbf{T}}_{pq}^{(j)^{\dagger} } \big]_{(v,v')} \bigg) \Bigg |^{2} \Bigg \}.\end{align}
\endgroup

\noindent Further, using Lemma~\ref{lem4}, i.e., $\Big [{\mathbf{T}}_{pq}^{(i)} {\mathbf{T}}_{pq}^{(i)^{\dagger} }\Big]_{(v,v^{\prime})} = 0$, and substituting $\mathbf{h}_{pq,i}=\boldsymbol{\varepsilon}_{pq,i}+\hat{\mathbf{h}}_{pq,i}$ yield
\begingroup\makeatletter\def\f@size{9}\check@mathfonts
\begin{align}&\hspace {-2pc}\mathbb {E} \big \{|\mathbb {I}_{qv1}|^{2}\big \}\nonumber\\
\stackrel {(a)}{=}&\sum _{p=1}^{N_{\mathrm{tx}}} \sum _{i=1}^{ L_{pq}} \sum _{j\neq i}^{ L_{pq}} {\eta }_{pq} {\kappa }_{pq}^{ij} \mathbb {E} \bigg \{ \Big | \boldsymbol{\varepsilon}_{pq,i} \hat{\mathbf{h}}_{pq,j}^* + \hat{\mathbf{h}}_{pq,i}\hat{\mathbf{h}}_{pq,j}^* \Big |^{2} \bigg \}\nonumber \\
\stackrel {(b)}{=}&\sum_{p=1}^{N_{\mathrm{tx}}}{\sum_{i=1}^{L_{pq}}{\sum_{j\ne i}^{L_{pq}}{\eta _{pq}}}}\kappa _{pq}^{ij}\mathrm{Tr\bigg(}\mathbb{E} \big\{\hat{\mathbf{h}}_{pq,j}^*\hat{\mathbf{h}}_{pq,j}^T\big\}\mathbb{E} \big\{\boldsymbol{\varepsilon }_{pq,i}^*\boldsymbol{\varepsilon }_{pq,i}^T\big\}\nonumber\\
&+\mathbb{E} \big\{\hat{\mathbf{h}}_{pq,j}^*\hat{\mathbf{h}}_{pq,j}^T\big\}\mathbb{E} \big\{\hat{\mathbf{h}}_{pq,i}^*\hat{\mathbf{h}}_{pq,i}^T\big\}\bigg)\nonumber\\
=&\sum _{p=1}^{N_{\mathrm{tx}}} \sum _{i=1}^{ L_{pq}} \sum _{j\neq i}^{ L_{pq}} {\eta }_{pq} {\kappa }_{pq}^{ij} \mathrm{Tr}\Big(\mathbf{R}_{pq,i}\mathbf{B}_{pq,j}\Big),\label{eq:Iq1}\end{align}
\endgroup

\noindent where in (a), the property that the variance of a sum of independent RVs is equal to the sum of the variances is used; (b) is based on the fact that $\boldsymbol{\varepsilon}_{pq,i}$ is zero-mean and independent of $\hat{\mathbf{h}}_{pq,i}$. Then, by applying Lemma~\ref{lem6}, and noting that users have zero-mean and independent channel gains, the inter-user interference term is obtained as follows:
\begingroup\makeatletter\def\f@size{9}\check@mathfonts
\begin{align} \mathbb {E} \big \{|\mathbb {I}_{qv2}|^{2}\big \}=&\mathbb {E} \left\{{ \left | \sum _{p=1}^{N_{\mathrm{tx}}} \sum _{\begin{subarray}{c}q^{\prime}\neq q\\q^{\prime}=0\end{subarray}}^{K_{\mathrm{u}}} \sum _{\substack {v^{\prime}=1 }}^{MN} {\eta }_{pq^{\prime} }^{1/2} [{\mathbf{H}}_{pq}]_{(v,:)} [\hat {\mathbf{H}}_{pq^{\prime}}^{\dagger}]_{(:,v^{\prime})} \right |^{2} }\right\}\nonumber \\
=&\sum _{p=1}^{N_{\mathrm{tx}}} \sum _{\begin{subarray}{c}q^{\prime}\neq q\\q^{\prime}=0\end{subarray}}^{K_{\mathrm{u}}} \sum _{i=1}^{ L_{pq}} \sum _{j=1}^{ L_{pq^{\prime}}} {\eta }_{pq^{\prime}} \mathrm{Tr}\Big(\mathbf{R}_{pq,i}\mathbf{B}_{pq^{\prime},j}\Big).\label{eq:Iq2}\end{align}
\endgroup

Finally, by substituting~\eqref{eq:DS},~\eqref{eq:BU},~\eqref{eq:Iq1},~and~\eqref{eq:Iq2} into~\eqref{eq:communication_SINR_ori}, the desired result in~\eqref{eq:DL_SE_old} is obtained following a series of algebraic manipulations.

\section{Proof of Corollary \ref{cor1}}
\label{app:cor1}

First, revisit Lemmas~\ref{lem4}~and~\ref{lem5}. If the paths between transmitting AP $p$ and user $q$ have different delay taps, that is, $\ell_{pq,i}\neq\ell_{pq,j}$, it can be verified that $\chi_{pq}^{ij}=0$ and $\kappa_{pq}^{ij}=1$, which consequently satisfies and establishes this lower bound.

Thus, only the paths between transmitting AP $p$ and user $q$ featuring the same delay tap, that is, $\ell_{pq,i}=\ell_{pq,j}$, remain to be examined. Without loss of generality, the first row of matrix ${\mathbf{T}}_{pq}^{(i)} {\mathbf{T}}_{pq'}^{(j)^{\dagger}}$ can be derived as~\cite[Appendix C]{mohammadi2022cell}
\begingroup\makeatletter\def\f@size{9}\check@mathfonts
\begin{align}&\hspace {-0.5pc} \Big[{\mathbf{T}}_{pq}^{(i)} {\mathbf{T}}_{pq'}^{(j)^{\dagger} }\Big]_{(1,:)} =\frac{1}{\sqrt{N}}\bigg [{\mathbf{a}}[{\mathbf{F}}_{N}^{\dagger}]_{(:,1)},\underbrace { 0, \ldots, 0}_{M-1}, {\mathbf{a}}[{\mathbf{F}}_{N}^{\dagger}]_{(:,2)}, \ldots,\nonumber \\&\qquad\qquad\qquad \qquad\qquad\qquad{\mathbf{a}}[{\mathbf{F}}_{N}^{\dagger}]_{(:,N)}, \underbrace { 0, \ldots, 0}_{M-1} \bigg],\end{align}
\endgroup

\noindent by denoting $z_{0}=z^{(k_{pq,i}-k_{pq,j})+(\kappa_{pq,i}-\kappa_{pq,j})}$, $\mathbf{a}$ is expressed as 
\begin{equation} {\mathbf{a}}= z_{0}^{- \ell _{pq,i}}\bigg [z_{0}^{MN}, z_{0}^{M}, \ldots, z_{0}^{M(N-1)} \bigg],\end{equation}

\noindent whose $k$th entry is $a_k$. Subsequently, $\chi_{pq}^{ij}$ and $\kappa_{pq}^{ij}$ can be respectively calculated by 
\begingroup\makeatletter\def\f@size{9}\check@mathfonts
\begin{align}\chi_{pq}^{ij}=\Bigg |\big [{\mathbf{T}}_{pq}^{(i)} {\mathbf{T}}_{pq}^{(j)^{\dagger} }\big]_{(1,1)}\Bigg |^{2} \!\!=\frac {1}{N^{2}} \Bigg |\sum _{k=0}^{N-1} a_{k} \Bigg |^{2}\!\! =\frac {1}{N^{2}}\! \sum _{k_{1}=0}^{N-1} \sum _{k_{2}=0}^{N-1} a_{k_{1}} a_{k_{2}}^{*},\end{align}
\endgroup

\begingroup\makeatletter\def\f@size{9}\check@mathfonts
\allowdisplaybreaks
\begin{align}\hspace {-1.8pc}\kappa_{pq}^{ij}=&\Bigg |\sum _{r'=2}^{MN} \big [{\mathbf{T}}_{pq}^{(i)} {\mathbf{T}}_{pq}^{(j)^{\dagger} }\big]_{(1,r')}\Bigg |^{2} =\frac {1}{N^{2}} \Bigg |\sum _{j=1}^{N-1}\sum _{k=0}^{N-1} a_{k} e^{j2\pi \frac {jk}{N}}\Bigg |^{2}\nonumber \\
=&\frac {1}{N^{2}} \sum _{k_{1}=0}^{N-1} \sum _{k_{2}=0}^{N-1} \sum _{j_{1}=1}^{N-1} \sum _{j_{2}=1}^{N-1} a_{k_{1}} a_{k_{2}}^{*} e^{-j2\frac {\pi }{N}(j_{1}-j_{2})(k_{1}-k_{2})}\nonumber \\
=&\frac {1}{N^{2}}\Bigg(N^{2} -\sum _{k_{1}=0}^{N-1}a_{k_{1}} \sum _{k_{2}=0}^{N-1}a_{k_{2}}^{*}\bigg( \sum _{j_{1}=0}^{N-1} e^{-j\frac {2\pi}{N}j_{1}(k_{1}-k_{2})}\nonumber\\
&\qquad\qquad+e^{j\frac {2\pi}{N}j_{1}(k_{1}-k_{2})}\bigg)+\sum _{k_{1}=0}^{N-1} \sum _{k_{2}=0}^{N-1} a_{k_{1}} a_{k_{2}}^{*}\Bigg)\nonumber \\
=&\frac {1}{N^{2}} \left(N^{2} -\sum _{k_{1}=0}^{N-1} a_{k_{1}} \sum _{k_{2}=0}^{N-1} a_{k_{2}}^{*}\Big(2N\delta (k_{1}-k_{2}) +1\Big) \right)\nonumber \\ 
=&\frac {1}{N^{2}} \left(-N^{2}+\sum _{k_{1}=0}^{N-1} a_{k_{1}} \sum _{k_{2}=0}^{N-1} a_{k_{2}}^{*}\right).\end{align}
\endgroup

Upon recognizing that $a_k$ is a complex number with unit modulus, it becomes evident that $\sum _{k_{1}=0}^{N-1} \sum _{k_{2}=0}^{N-1} a_{k_{1}} a_{k_{2}}^{*}\leq N^{2}$, which implies $\chi_{pq}^{ij}+\kappa_{pq}^{ij}\leq1$. Thus, by setting $\chi_{pq}^{ij}+\kappa_{pq}^{ij}=1$ in the denominator of \eqref{eq:DL_SE_old} defines \eqref{eq:DL_SE} as a lower bound.

\section{Proof of Lemma \ref{lem2}}
\label{app:lem2}

The sensing SINR for the sensing received signal in~\eqref{eq:sensing_received} is defined by
\begingroup\makeatletter\def\f@size{9}\check@mathfonts
\begin{align}
\text{SINR}&^{(\mathrm s)}=\nonumber\\
&\frac{\rho_{\mathrm{d}}\sum\limits_{r=1}^{N_{\mathrm{rx}}}\mathbb{E}\left\{\Big\| \sum\limits_{p=1}^{N_{\mathrm{tx}}}\sum\limits_{q=0}^{K_{\mathrm{u}}}\eta _{pq}^{1/2}{\mathbf{H}}_{pr} \hat {\mathbf{H}}_{pq}^{\dagger}\mathbf{s}_{q} \Big\|^2\right\}}{\rho_{\mathrm{d}}\sum\limits_{r=1}^{N_{\mathrm{rx}}}\mathbb{E}\left\{\Big\| \sum\limits_{p=1}^{N_{\mathrm{tx}}}\sum\limits_{q=0}^{K_{\mathrm{u}}}\eta _{pq}^{1/2}{\mathbf{G}}_{pr} \hat {\mathbf{H}}_{pq}^{\dagger}\mathbf{s}_{q} \Big\|^2\right\}+\sum\limits_{r=1}^{N_{\mathrm{rx}}}\mathbb{E}\Big\{\|\mathbf{w}_r\|^2\Big\}}.
\label{eq:sensing_SINR_ori}
\end{align}
\endgroup

\noindent The expectation in the numerator can be computed as
\begingroup\makeatletter\def\f@size{9}\check@mathfonts
\begin{align}
\rho_{\mathrm{d}}\sum_{r=1}^{N_{\mathrm{rx}}}&\mathbb{E}\left\{\bigg\| \sum_{p=1}^{N_{\mathrm{tx}}}\sum_{q=0}^{K_{\mathrm{u}}}\eta _{pq}^{\frac12}{\mathbf{H}}_{pr} \hat {\mathbf{H}}_{pq}^{\dagger}\mathbf{s}_{q} \bigg\|^2\right\}\nonumber\\
=&\rho_{\mathrm{d}}\sum_{v=1}^{MN}\sum_{r=1}^{N_{\mathrm{rx}}}\mathbb{E}\Bigg\{\bigg\| \sum_{p=1}^{N_{\mathrm{tx}}}\sum_{q=0}^{K_{\mathrm{u}}}\sum_{v^{\prime}=1}^{MN}\eta _{pq}^{\frac12}\sigma_{pr}\beta_{pr}^{\frac12}\nonumber \\
&\!\!\times\left( \mathbf{h}_{tr}^{\phantom{T}}\mathbf{h}_{pt}^{T}\otimes \left[\mathbf{T}_{pr}\right]_{(v,:)} \right) \left(\hat{\mathbf h}_{pq,i}^*\otimes \big[{\mathbf{T}}_{pq}^{(i)}\big]_{(:,v^{\prime})}\right){s}_{qv^{\prime}} \bigg\|^2\Bigg\}\nonumber\\
\stackrel {(a)}{=}&\rho_{\mathrm{d}}\sum_{v=1}^{MN}\sum_{r=1}^{N_{\mathrm{rx}}}\mathbb{E}\Bigg\{\bigg\| \sum_{p=1}^{N_{\mathrm{tx}}}\sum_{q=0}^{K_{\mathrm{u}}}\sum_{v^{\prime}=1}^{MN}\eta _{pq}^{\frac12}\sigma_{pr}\beta_{pr}^{\frac12}\sum_{i=1}^{L_{pq}}\mathbf{h}_{tr}\mathbf{h}_{pt}^{T}\hat{\mathbf h}_{pq,i}^*\nonumber\\
&\times\big[\mathbf{T}_{pr}\mathbf{T}_{pq}^{(i)}\big]_{(v,v^{\prime})}{s}_{qv^{\prime}}\bigg\|^2\Bigg\}\nonumber\\
\stackrel {(b)}{=}&\rho_{\mathrm{d}}MN\sum_{r=1}^{N_{\mathrm{rx}}} \sum_{p=1}^{N_{\mathrm{tx}}}\sum_{q=0}^{K_{\mathrm{u}}}\sum_{i=1}^{L_{pq}}\eta _{pq}\sigma_{pr}^2\beta_{pr}\mathbb{E}\left\{\left\|\mathbf{h}_{tr}^{\phantom{T}}\mathbf{h}_{pt}^{T}\hat{\mathbf h}_{pq,i}^*\right\|^2\right\}\nonumber\\
=&\rho_{\mathrm{d}}MN\sum_{r=1}^{N_{\mathrm{rx}}} \sum_{p=1}^{N_{\mathrm{tx}}}\sum_{q=0}^{K_{\mathrm{u}}}\sum_{i=1}^{L_{pq}}\eta _{pq}\sigma_{pr}^2\beta_{pr}\mathrm{Tr}\left( \mathbf{V}_{pr}^{\phantom{\dagger}}\mathbf{B}_{pq,i}\mathbf{V}_{pr}^{\dagger} \right),
\label{eq:sensing_num}
\end{align}
\endgroup

\noindent where (a) follows from the identity $(\mathbf{A}\otimes\mathbf{B})(\mathbf{C}\otimes\mathbf{D})=(\mathbf{AC}\otimes\mathbf{BD})$, and (b) holds based on the assumptions that the RCSs from different transmitting APs are independent, that is, $\operatorname{cov}\left(\alpha_{pr},\alpha_{p^{\prime}r}\right)=0$ and ${s}_{qv^{\prime}}$ denotes the i.i.d. RV.

By following similar steps for deriving the numerator of~\eqref{eq:sensing_SINR_ori}, the expectation in the denominator of the sensing SINR in~\eqref{eq:sensing_SINR_ori} can be expressed by
\begingroup\makeatletter\def\f@size{9}\check@mathfonts
\allowdisplaybreaks
\begin{align}
\rho_{\mathrm{d}}\sum\limits_{r=1}^{N_{\mathrm{rx}}}&\mathbb{E}\left\{\Big\| \sum\limits_{p=1}^{N_{\mathrm{tx}}}\sum\limits_{q=0}^{K_{\mathrm{u}}}\eta _{pq}^{1/2}{\mathbf{G}}_{pr} \hat {\mathbf{H}}_{pq}^{\dagger}\mathbf{s}_{q} \Big\|^2\right\}\nonumber\\
=&\rho_{\mathrm{d}}MN\sum_{r=1}^{N_{\mathrm{rx}}} \sum_{p=1}^{N_{\mathrm{tx}}}\sum_{q=0}^{K_{\mathrm{u}}}\sum_{i=1}^{L_{pq}}\eta _{pq}\mathbb{E}\bigg\{\mathrm{Tr}\bigg(\mathbf{R}_{\mathrm{rx},(pr)}^{\frac12}\mathbf{W}_{\mathrm{ch},(pr)}\nonumber\\
&\times\Big(\mathbf{R}_{\mathrm{tx},(pr)}^{\frac12}\Big)^T \mathbf{B}_{pq,i} \Big(\mathbf{R}_{\mathrm{tx},(pr)}^{\frac12}\Big)^*\mathbf{W}_{\mathrm{ch},(pr)}^{\dagger}\Big(\mathbf{R}_{\mathrm{rx},(pr)}^{\frac12}\Big)^{\dagger}\bigg)\bigg\}\nonumber\\
\stackrel {(a)}{=}&\rho_{\mathrm{d}}MN\sum_{r=1}^{N_{\mathrm{rx}}} \sum_{p=1}^{N_{\mathrm{tx}}}\sum_{q=0}^{K_{\mathrm{u}}}\sum_{i=1}^{L_{pq}}\eta _{pq}\mathrm{Tr}\bigg(\mathbb{E}\Big\{\mathbf{W}_{\mathrm{ch},(pr)}^{\dagger}\mathbf{R}_{\mathrm{rx},(pr)}\nonumber\\
&\qquad\qquad\ \times\mathbf{W}_{\mathrm{ch},(pr)}\Big\}\Big(\mathbf{R}_{\mathrm{tx},(pr)}^{\frac12}\Big)^T\mathbf{B}_{pq,i}\Big(\mathbf{R}_{\mathrm{tx},(pr)}^{\frac12}\Big)^*\bigg)\nonumber\\
\stackrel {(b)}{=}&\rho_{\mathrm{d}}MN\sum\limits_{r=1}^{N_{\mathrm{rx}}}{\sum\limits_{p=1}^{N_{\mathrm{tx}}}{\sum\limits_{q=0}^{K_{\mathrm{u}}}{\sum\limits_{i=1}^{L_{pq}}{\eta _{pq}\mathrm{Tr}\left( \mathbf{R}_{\mathrm{rx},(pr)} \right)\mathrm{Tr}\left( \mathbf{R}_{\mathrm{tx},(pr)}\mathbf{B}_{pq,i} \right) }}}},
\end{align}
\endgroup

\noindent where (a) follows from the property $\operatorname{tr}(\mathbf{AB})=\operatorname{tr}(\mathbf{BA})$ and the fact that $\mathbf{R}_{\mathrm{rx},(pr)}^{1/2}$ is Hermitian symmetric; (b) is based on the lemma defining that $\mathbb{E}\left\{\mathbf{W}_{\mathrm{ch},(pr)}^{\dagger}\mathbf{R}_{\mathrm{rx},(pr)}\mathbf{W}_{\mathrm{ch},(pr)}\right\}=\mathrm{tr}\left(\mathbf{R}_{\mathrm{rx},(pr)}\right)\mathbf{I}_{M_\mathrm{t}}$ in~\cite{behdad2024multi}. Then, by expanding $\mathbb{E}\left\{\|\mathbf{w}_r\|^2\right\}=M_\mathrm{t}MN$ and substituting \eqref{eq:sensing_SINR_ori}, the sensing SINR can be obtained by~\eqref{eq:sensing_SINR}.

\section{Proof of Lemma \ref{lem3}}
\label{app:lem3}

Invoking \eqref{eq:x_p} yields 
\begingroup\makeatletter\def\f@size{9}\check@mathfonts
\begin{align}
	\mathbb{E} &\left\{ \| \mathbf{x}_{p}\| ^2 \right\}\nonumber\\
	&=\mathbb{E} \left\{ \left\| \sqrt{\rho_{\mathrm{d}}}\sum_{q=1}^{K_{\mathrm{u}}}{\eta _{pq}^{1/2}}\hat{\mathbf{H}}_{pq}^{\dagger}\mathbf{s}_q\right\| ^2 \right\}\nonumber\\
	&=\rho_{\mathrm{d}}\mathbb{E} \left\{ \mathrm{Tr}\left( \left[ \sum_{q=1}^{K_{\mathrm{u}}}{\eta _{pq}^{1/2}}\hat{\mathbf{H}}_{pq}^{\dagger}\mathbf{s}_q \right] \left[ \sum_{q^{\prime}=1}^{K_{\mathrm{u}}}{\eta _{pq^{\prime}}^{1/2}}\mathbf{s}_{q^{\prime}}^{\dagger}\hat{\mathbf{H}}_{pq^{\prime}} \right] \right) \right\}\nonumber\\
	&\stackrel {(a)}{=}\rho_{\mathrm{d}}\sum_{q=1}^{K_{\mathrm{u}}}{\eta _{pq}}\mathrm{Tr}\left( \mathbb{E} \{\hat{\mathbf{H}}_{pq}^{\dagger}\hat{\mathbf{H}}_{pq}\} \right)\nonumber\\
	&=\rho_{\mathrm{d}}\sum_{q=1}^{K_{\mathrm{u}}}{\eta _{pq}\mathrm{Tr}\!\!\left( \!\mathbb{E} \left\{\sum_{i=1}^{L_{pq}}{\left( \mathbf{h}_{pq,i}^{*}\otimes \mathbf{T}_{pq}^{(i)\dagger} \right)\!\! \left( \mathbf{h}_{pq,i}^{T}\otimes \mathbf{T}_{pq}^{(i)} \right)}\right\} \right)}\nonumber\\
	&=\rho_{\mathrm{d}}\sum_{q=1}^{K_{\mathrm{u}}}{\sum_{i=1}^{L_{pq}}{\eta _{pq}}\mathrm{Tr}\left(\mathbf{B}_{pq,i}\otimes \mathbf{I}_{MN}\right)}\nonumber\\
	&=\rho_{\mathrm{d}}MN\sum_{q=1}^{K_{\mathrm{u}}}{\sum_{i=1}^{L_{pq}}{\eta _{pq}}\mathrm{Tr}\left(\mathbf{B}_{pq,i}\right)} \, ,
\end{align}
\endgroup

\noindent where in (a), we assume $\mathbf{s}_{q}$ has i.i.d. entries. Based on~\eqref{eq:DD_channel} and the fact that different paths have zero-mean i.i.d channel gains, that is, $\mathbb{E} \{\hat{\mathbf{h}}_{pq,i}^*\hat{\mathbf{h}}_{pq,j}^T\}=\mathbf{0}_{M_{\mathrm{t}}\times M_{\mathrm{t}}}, \forall i\neq j$, and using Lemma~\ref{lem4}, the derivation is completed.

}

\bibliographystyle{IEEEtran}
\bibliography{IEEEabrv,bibRef}

\end{document}